\documentclass[journal]{IEEEtran}
\IEEEoverridecommandlockouts
\usepackage{cite}
\usepackage{amsmath,amssymb,amsfonts}
\usepackage{algorithmic}
\usepackage{graphicx}
\usepackage[skip=2pt]{caption}
\usepackage{url}
\usepackage{subcaption}  
\usepackage{textcomp}
\usepackage{xcolor}
\begin{document}

\title{Lifecycle Cost-Effectiveness Modeling for Redundancy-Enhanced Multi-Chiplet Architectures \\
}
\author{
    Zizhen~Liu,~\IEEEmembership{Member,~IEEE,}
    Fangzhiyi~Wang,
    Mengdi~Wang,~\IEEEmembership{Member,~IEEE,}
    Jing~Ye,~\IEEEmembership{Member,~IEEE,}
    Hayden~Kwok-Hay~So,~\IEEEmembership{Senior Member,~IEEE,}
    Cheng~Liu,~\IEEEmembership{Senior Member,~IEEE,}\\
    and Huawei~Li,~\IEEEmembership{Senior Member,~IEEE}

\thanks{This paper is supported in part by National Natural Science Foundation of China (NSFC) under grant No. 92373206 and 62174162 and in part by the Strategic Priority Research Program of the Chinese Academy of Sciences, Grant No. XDB0660102}
\thanks{Z. Liu and M. Wang are with the State Key Lab of Processors, Institute of Computing Technology, Chinese Academy of Sciences. E-mail: \{liuzizhen, wangmengdi\}@ict.ac.cn.}

\thanks{Z. Liu and M. Wang are with the State Key Lab of Processors, Institute of Computing Technology, Chinese Academy of Sciences. E-mail: \{liuzizhen, wangmengdi\}@ict.ac.cn.}
\thanks{F. Wang and C. Liu are with the State Key Lab of Processors, Institute of Computing Technology, Chinese Academy of Sciences, and the University of Chinese Academy of Sciences. E-mail: 18225532058@163.com; liucheng@ict.ac.cn.}
\thanks{J. Ye and H. Li are with the State Key Lab of Processors, Institute of Computing Technology, Chinese Academy of Sciences, the University of Chinese Academy of Sciences, and CASTEST Co., Ltd. E-mail: \{yejing, lihuawei\}@ict.ac.cn.}
\thanks{H. K.-H. So is with the School of Innovation, The University of Hong Kong. E-mail: hso@eee.hku.hk.}
}

\markboth{IEEE TRANSACTIONS ON COMPUTER-AIDED DESIGN OF INTEGRATED CIRCUITS AND SYSTEMS, ~Vol.~14, No.~8, August~2015}%
{Shell \MakeLowercase{\textit{et al.}}: Lifecycle Cost-Effectiveness Modeling for Redundancy-Enhanced Multi-Chiplet Architectures }

\maketitle
\begin{abstract}
The growing demand for compute-intensive applications has made multi-chiplet architectures a promising alternative to monolithic designs, offering improved scalability and manufacturing flexibility. However, effectively managing the economic effectiveness remains challenging. Existing cost models either overlook the amortization of compute value over a chip’s operational lifetime or fail to evaluate how redundancy strategies—which are widely adopted to enhance yield and fault tolerance—impact long-term cost efficiency.

This paper presents a comprehensive cost-effectiveness framework for multi-chiplet architectures, introducing a novel Lifecycle Cost Effectiveness (LCE) metric that evaluates amortized compute costs by jointly optimizing manufacturing expenses and operational lifetime. Our approach uniquely integrates: (1) redundancy-aware cost modeling spanning both intra- and inter-chiplet levels, (2) reliability-driven lifetime estimation, and (3) quantitative analysis of how redundancy configurations on overall economic effectiveness. Extensive trade-off and multi-objective optimization studies demonstrate the effectiveness of the model and reveal essential co-optimization strategies between module and chiplet-level redundancy to achieve cost-efficient multi-chiplet architecture designs.  
\end{abstract}


\section{Introduction}
\IEEEPARstart{T}{he} rapid proliferation of AI—particularly large language models (LLMs) and generative AI—has driven an unprecedented surge in computational demands\cite{lu2024blending}. While a natural response is to develop large-scale monolithic AI chips, this approach faces significant limitations. As transistor scaling approaches its physical limits, increasing the die area to integrate more transistors leads to reduced wafer utilization and greater susceptibility to manufacturing defects, ultimately resulting in lower yields and higher production costs. To overcome these challenges, multi-chiplet architectures integrated via advanced 2.5D/3D packaging has emerged as a compelling alternative, offering both improved scalability and manufacturing efficiency\cite{naffziger2021pioneering}. At the same time, the rapidly escalating cost of computing infrastructure has become a major barrier to the widespread deployment of advanced AI models. Multi-chiplet integration not only meets the performance requirements of modern AI models but also enables more cost-effective and flexible design, making it a crucial strategy for future AI hardware development.

Despite the advantages of chiplet-based design, such as enhanced
manufacturing flexibility and potential die-level yield improvements,
integrating many chiplets in a high-density 2.5D/3D package introduces
new challenges. First, assembly-induced defects, including microbump
failures, non-functional TSVs, and interposer-level variations, can 
significantly reduce the yield of the assembled multi-chiplet system\cite{wu2024advances}. 
Second, as the scale of chip integration grows, the reliability of the entire multi-chiplet architecture increasingly depends on its weakest component: The failure of a single chiplet can adversely affect the functionality of the entire chip, potentially leading to significant reductions in its operational lifetime. Additionally, unlike general-purpose computing workloads, which may vary in computational and memory demands, AI workloads are typically both compute- and memory-intensive, exacerbating depreciation and operational costs.

To address these risks, incorporating redundancy at both the intra-chiplet and inter-chiplet levels has become a key design strategy\cite{gehle2024reliability}. While redundancy
also exists in monolithic SoCs, it is primarily applied within a single die. In contrast,
the redundancy mechanisms considered in this work operate across chiplets,
including module-level redundancy within each chiplet, routing-level redundancy
in the chiplet Network-on-Package, and inter-chiplet redundancy that improves yield during
chiplet-to-interposer integration. Although redundancy can enhance chip yield and reliability, it also introduces complex trade-offs involving area overhead, manufacturing costs, and long-term depreciation of computational resources. Notably, redundancy directly affects the lifetime, which in turn influences the total computational output over the chip's operational lifespan and the amortized cost per unit of compute\cite{li2020chiplet}. The extent to which redundancy improves economic efficiency depends not only on the redundancy strategy itself, but also on how these design choices interact with reliability and operational depreciation over time. Thus, effectively quantifying and optimizing these trade-offs is critical for designing economically viable multi-chiplet architectures\cite{ahmad2022heterogeneous}.

Although there have been extensive studies of cost modeling and optimization of multi-chiplet architectures\cite{stow2016cost,stow2017cost,jerger2014noc,feng2022chiplet,ahmad2022heterogeneous}, existing studies rarely offer cost models specifically tailored to redundancy strategies, leaving the quantitative evaluation of their trade-offs largely unexplored. Moreover, most current chiplet cost models focus primarily on  manufacturing costs, often overlooking the depreciation of compute value over the chip’s operational lifetime\cite{feng2022chiplet, stow2016cost,stow2017cost,sahoo2024trade}. Relying solely on absolute engineering cost as a metric is inadequate, as it fails to account for critical factors such as reliability, lifetime, and the nonlinear scaling relationship between compute capacity and total cost. This limits the applicability of such models in guiding economically optimal chiplet-based architecture design.

To bridge these gaps, we propose a cost-effectiveness modeling framework for multi-chiplet architectures that quantitatively evaluates how redundancy strategies at both intra-chiplet and inter-chiplet levels affect overall economic efficiency. Unlike traditional models that consider only manufacturing-related non-recurring engineering (NRE) and recurring engineering (RE) costs, our framework introduces a Lifecycle Cost Effectiveness (LCE) metric. This metric captures the amortized cost per unit of compute output by relating total NRE and RE costs to the compute capacity delivered over the chip’s operational lifetime. Compute capacity, in this context, reflects both spatial resources (e.g., number of active chiplets) and temporal factors (e.g., chip longevity). As a result, the LCE-based model provides a comprehensive and standardized evaluation of unit cost benefit, enabling more informed design decisions under both performance and economic constraints.

The primary contributions of this work are as follows:

\begin{itemize}

    \item 
        We present the first cost-effectiveness modeling framework for multi-chiplet architectures that explicitly accounts for reliability-driven lifetime variation generally overlooked by prior models.

    \item
        We propose the Lifecycle Cost Effectiveness (LCE) metric, which
        integrates manufacturing cost, yield, architecture degradation,
        and lifetime-delivered compute into a unified formulation.
        This enables amortized cost per compute unit to be evaluated
        over the entire operational lifetime.

    \item 
        We provide the first quantitative analysis of how module-level,
        routing-level, and inter-chiplet redundancy strategies influence
        lifetime, yield, and long-term economic efficiency of
        multi-chiplet architectures.


\end{itemize}

\section{ Background and Related Work}
\subsection{Chiplet Lifecycle and Fault Tolerance Background}

Chiplets are pre-manufactured dies with specific functions that are integrated into larger systems through advanced packaging. In multi-chiplet architectures, multiple dies are attached to an interposer via micro-bumps to form a high-density interconnection network, which then interfaces with the package substrate and external systems~\cite{ma2022survey}. This modular integration improves manufacturing flexibility, scalability, and die yield, but also introduces additional area overhead, packaging complexity, and new dimensions in system-level cost and fault tolerance modeling~\cite{taheri2022deft}.

\begin{figure}[!t]
    \centering
    \includegraphics[width=0.49\textwidth]{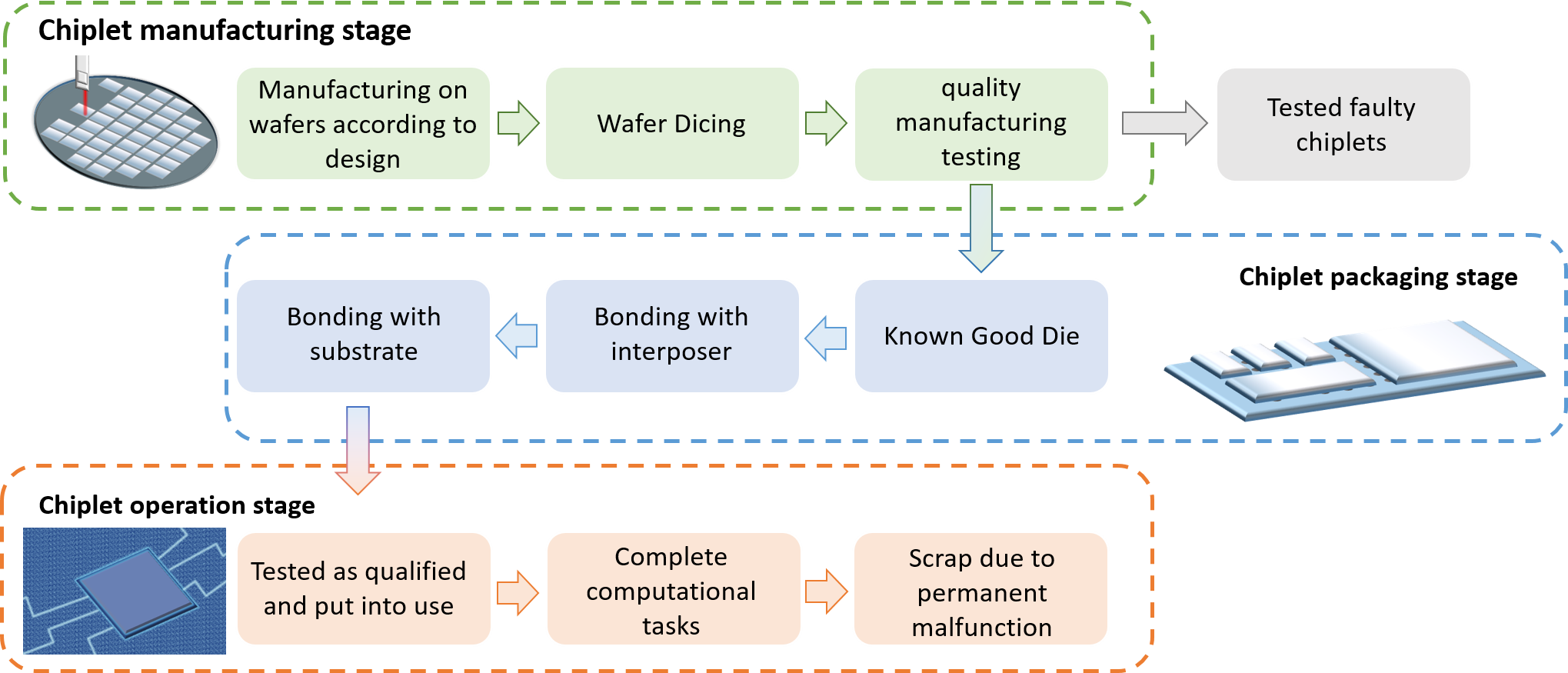}
    \captionsetup{skip=0pt}
    \caption{Integrated Chip Manufacturing and Operation Lifecycle.}
    \label{fig:background}
\end{figure}

As shown in Fig~\ref{fig:background}, the lifecycle of chiplet-based systems consists of three major stages.
(1) Manufacturing: Chiplets are fabricated on wafers, diced, and tested. Only known-good dies (KGDs) proceed to assembly to avoid yield loss.
(2) Packaging: Verified chiplets are bonded to the interposer and the package substrate, forming a complete system-in-package that undergoes final testing.
(3) Operation: The deployed system executes real-world workloads over its lifetime. Aging, wear-out, and environmental disturbances gradually degrade components, potentially leading to permanent functional faults. Depending on the severity and the availability of mitigation mechanisms, the system may operate with degraded performance or be retired once functionality becomes unacceptable.

Permanent faults may originate from both die fabrication and packaging-level processes such as substrate manufacturing, interposer fabrication, and bonding. Although chiplet partitioning improves die-level yield, interposer-based integration introduces additional defect opportunities and can reduce package-level yield due to more complex assembly steps. Runtime aging and environmental stress further amplify the risk of permanent failures during operation~\cite{ma2023review}.

To enhance robustness, chiplet systems frequently incorporate hardware redundancy—such as spare cores, spare wires, or redundant interconnect links—to improve initial yield and extend system lifespan~\cite{shamshiri2008cost, shamshiri2009yield}. Redundant components can be activated to bypass defective structures as failures accumulate, providing significant benefits in both manufacturing cost reduction and runtime reliability. However, redundancy also introduces area and power overheads, making excessive redundancy economically undesirable~\cite{agrawal2023level, huang2010characterizing}.

Therefore, evaluating fault-tolerant chiplet systems requires more than assessing upfront manufacturing cost. The computational value delivered across the entire operational lifetime, including performance throughput, resilience against permanent faults, and effective lifespan, must also be considered. This motivates the need for a lifecycle-aware cost efficiency metric that jointly captures compute capability, fault-tolerant longevity, and total cost of ownership.

\subsection{Related Work}
Yield and cost modeling have long played a central role in semiconductor manufacturing analysis. Early works such as Cunningham~\cite{cunningham1990use} established foundational wafer-level yield models, while Lynch~\cite{lynch1977reduction} and Gwennap~\cite{gwennap1993estimating} investigated alignment yield and die-size-based cost estimation. With the emergence of chiplet architectures, these classical models have been extended to capture cost components unique to multi-die integration. Stow et al.~\cite{stow2016cost} examined engineering costs in 2.5D/3D integration, and Mercier et al.~\cite{mercier2006yield} analyzed cost trade-offs between active/passive interposers and 3D stacking. Feng et al.~\cite{feng2022chiplet} incorporated die-to-die interconnect overhead into NRE modeling, while Ehrett et al.~\cite{ehrett2021chopin} studied chip reuse and design decomposition. Tang and Xie~\cite{tang2022costaware} developed a framework for 2.5D cost estimation across interconnect types and technology nodes, and Gopalakrishnan et al.~\cite{gopalakrishnan2011process} evaluated process-level costs including labor and equipment.

More recent studies focus on system-level and heterogeneous chiplet integration. Ahmad et al.~\cite{ahmad2022heterogeneous} modeled SoCs as multi-chip assemblies and provided detailed cost breakdowns across materials, testing, KGD strategies, and operational factors. Chen et al.~\cite{chen2023floorplet} introduced \textit{Floorplet}, integrating performance modeling with cost and reliability optimization, while Ning et al.~\cite{ning2023supply} characterized the entire manufacturing supply chain and proposed design methodologies to reduce cost and time-to-market. Other recent efforts emphasize PPAC-driven multi-die floorplanning~\cite{roman2025ppac}, chiplet dimension and packaging trade-offs~\cite{sahoo2024trade}, and cost-aware spatial accelerator specialization (\textit{Monad})~\cite{hao2023monad}. Yang et al.~\cite{yang2024challenges} and Mallya et al.~\cite{mallya2025performance} evaluated challenges, benefits, and performance overheads of chiplet systems for AI computing. Cost-driven tools such as \textit{CATCH}~\cite{graening2025catch} and \textit{ChipletPart}~\cite{graening2025chipletpart} further enable co-optimization of heterogeneous chiplet layouts, while Ayes et al.~\cite{ayes2025network} explored cost–performance co-design for network-on-interposer architectures. Mishty et al.~\cite{mishty2024ai} demonstrated AI-aided system and design co-optimization for chiplet-based AI accelerators using emerging memories and advanced packaging.

Fault tolerance and redundancy have also been studied, though mainly from a manufacturing or yield-improvement perspective. Lauterbach~\cite{lauterbach2021path} discussed wafer-scale redundancy strategies, whereas Shamshiri et al.~\cite{shamshiri2011modeling} and Ahmad et al.~\cite{ahmad2022heterogeneous} analyzed SoC and chiplet redundancy for yield enhancement. Xu et al.~\cite{xu2010tsv} highlighted TSV overheads in 3D NoCs, underscoring the cost impact of interconnect reliability. However, existing approaches generally lack quantitative frameworks for trading off redundancy cost, operational reliability, and lifetime benefits in multi-chiplet architectures.

In summary, existing works provide foundational cost and yield modeling
approaches; however, they focus primarily on manufacturing-side cost factors.
The broader economic implications—such as how reliability, lifetime, and
redundancy influence compute value over time—remain insufficiently explored.

\section{Motivation}

\begin{figure}[htbp]
    \centering

    \begin{subfigure}[b]{0.48\textwidth}  
        \includegraphics[width=\textwidth]{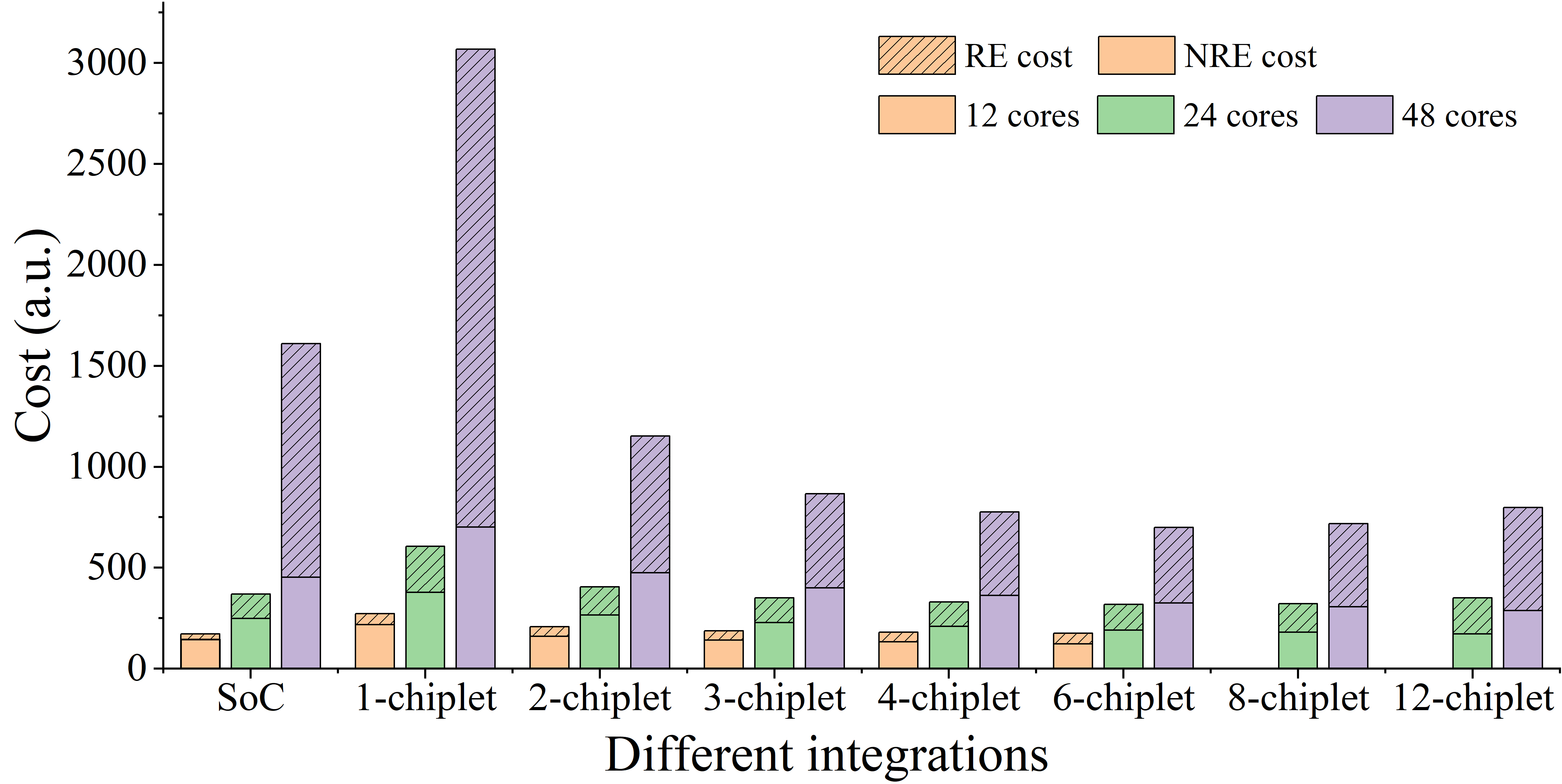}
        \caption{Cost Variation with Different Chiplet Integration Strategies.}
        \label{fig:sub1}
    \end{subfigure}
    \hfill  
    \begin{subfigure}[b]{0.48\textwidth}
        \includegraphics[width=\textwidth]{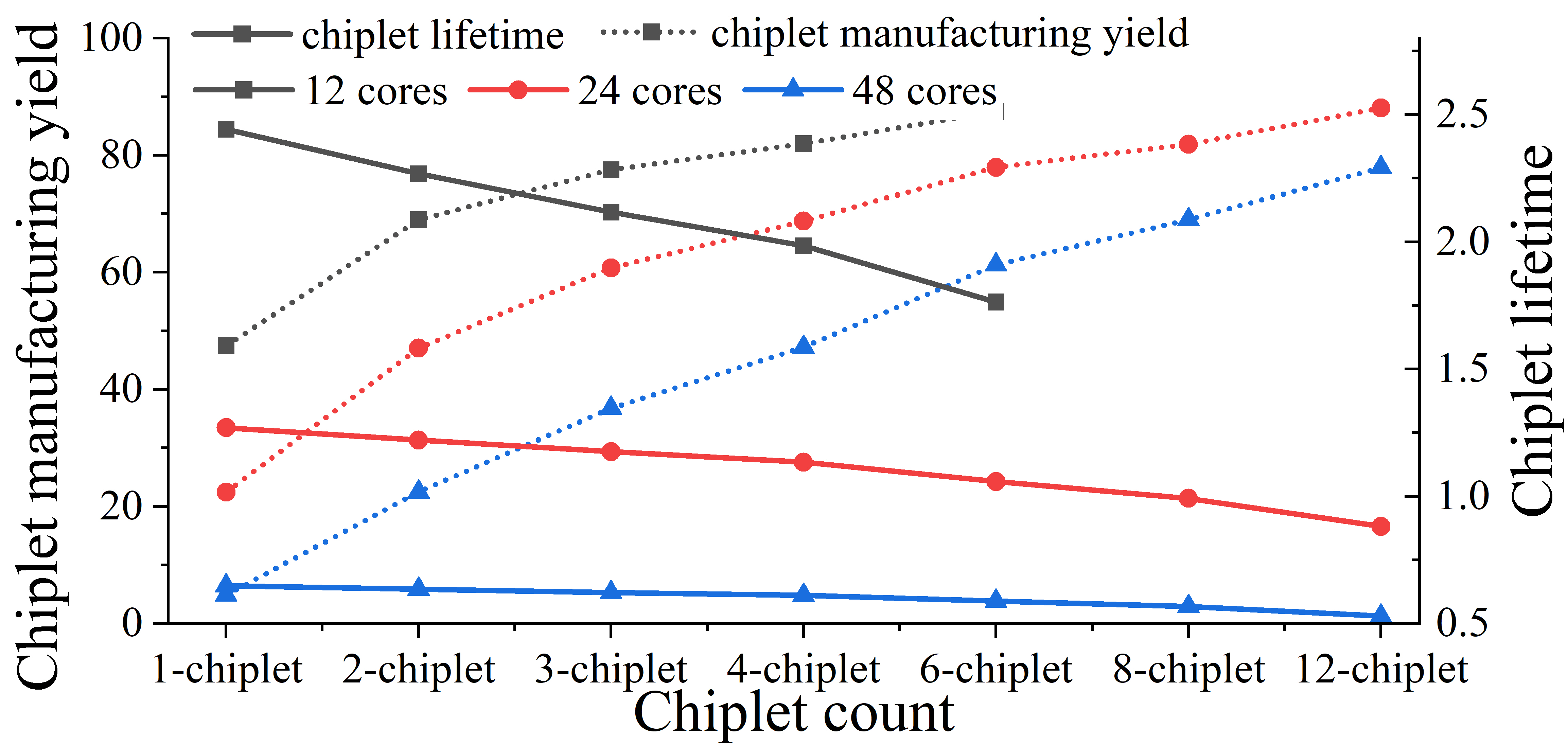}
        \caption{Yield and Lifetime Variation with the Number of Chiplets.}
        \label{fig:sub2}
    \end{subfigure}

    \caption{The configuration optimized for chiplet manufacturing cost can be suboptimal for lifecycle cost-effectiveness, as different chiplet arrangements can affect key factors such as operational lifetime and manufacturing yield differently across the overall multi-chiplet architecture.}
    \label{fig:framework}
\end{figure}

While prior cost models characterize manufacturing and integration expenses,
they do not capture the nonlinear relationship among compute capacity, total
cost, and lifetime. Moreover, they overlook how reliability-induced lifetime
variation and redundancy strategies reshape the effective cost of delivered
compute. These gaps motivate the need for a lifecycle-aware cost-effectiveness
framework.                   

As illustrated in Fig.~\ref{fig:sub1}, increasing core counts from 24 to 48 in monolithic systems results in a cost escalation of over 300\%, primarily due to yield degradation and die scaling limitations. In contrast, scaling to six chiplets within a multi-chiplet package increases cost by only 100\%, owing to improved yield and modularity. This disparity highlights that absolute manufacturing cost alone does not accurately reflect the cost per unit of computational capacity, which is critical for assessing architectural cost-effectiveness under practical scaling scenarios.


Furthermore, A system's usable lifetime directly affects how engineering costs are amortized
into effective cost per compute over time. A longer lifetime improves cost amortization and reduces the annualized cost per compute unit. However, as shown in Fig.~\ref{fig:sub2}, although chiplet-based integration improves manufacturing yield, it introduces added packaging and interconnect complexity that compromises system-level reliability. This is reflected by a decline in Mean Time To Failure (MTTF)\cite{wasala2023lifetime}, shortening the usable lifetime and diminishing the depreciation efficiency of compute capacity over time. Consequently, the effective cost per usable compute unit increases, especially in high-reliability or long-lifetime deployment scenarios\cite{sapra2023exploring}.


Crucially, existing models often overlook the complex interdependencies among reliability—which directly impacts manufacturing yield and operational lifetime—and the resulting effects on cost amortization. Consequently, they fail to accurately quantify how redundancy and other fault-tolerance strategies affect the effective amortized cost of compute capacity over the system’s lifetime. This limitation hinders a comprehensive evaluation of the true economic trade-offs of redundancy across diverse workload and reliability requirements.

Motivated by these limitations, we propose a redundancy-aware cost-effectiveness modeling framework for multi-chiplet architectures. This framework jointly incorporates engineering costs, reliability-induced lifetime variations, and amortization of compute capacity over the lifecycle, enabling principled optimization of economic efficiency under diverse redundancy strategies and compute workloads.

\section{Chiplet Cost-effectiveness Model}

\subsection{Overview of The Proposed Lifecycle Cost-effectiveness Model}
In this section, we present a unified cost-effectiveness modeling framework for multi-chiplet architectures with redundancy enhancement. Considering the inherent complexity of practical architectures, certain secondary factors are either simplified or omitted in this model for tractability. The primary assumptions are outlined as follows.

\begin{figure}[htbp]
    \centering
    \begin{subfigure}[b]{0.45\linewidth}
        \centering
        \includegraphics[width=\linewidth]{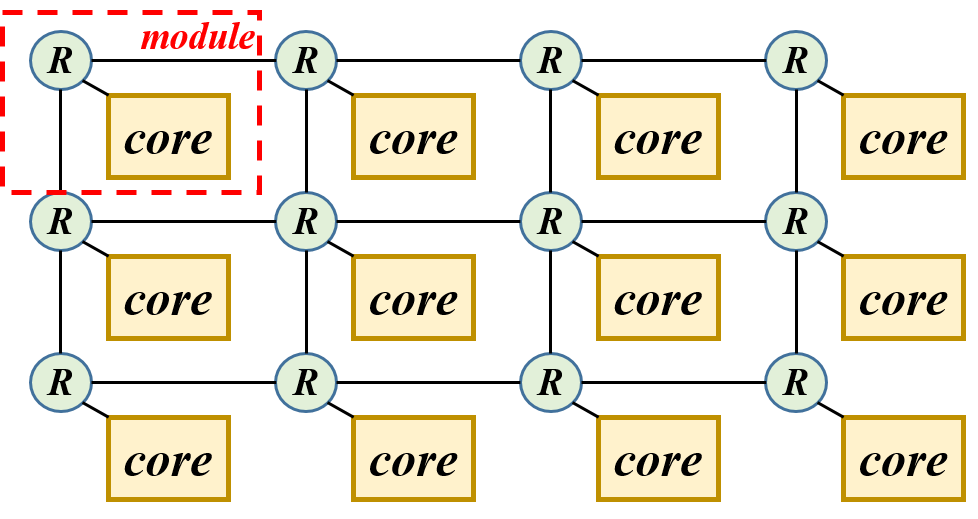}
        \caption{The single chiplet model.}
        \label{fig2:sub1}
    \end{subfigure}
    \begin{subfigure}[b]{0.45\linewidth}
        \centering
        \includegraphics[width=\linewidth]{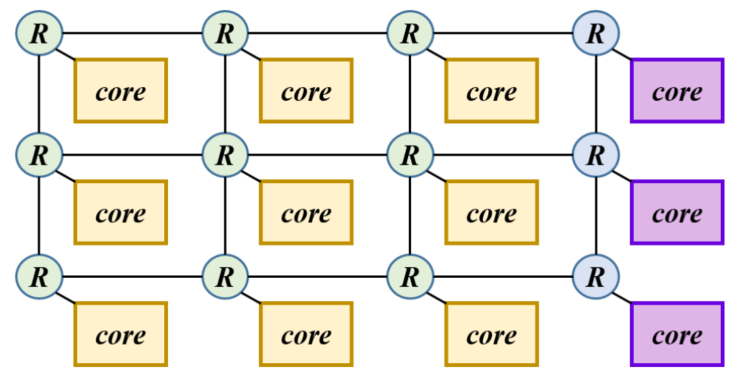}
        \caption{Module redundancy.}
        \label{fig2:sub2}
    \end{subfigure}
    
    \vspace{0.3cm} 
    
    \begin{subfigure}[b]{0.45\linewidth}
        \centering
        \includegraphics[width=\linewidth]{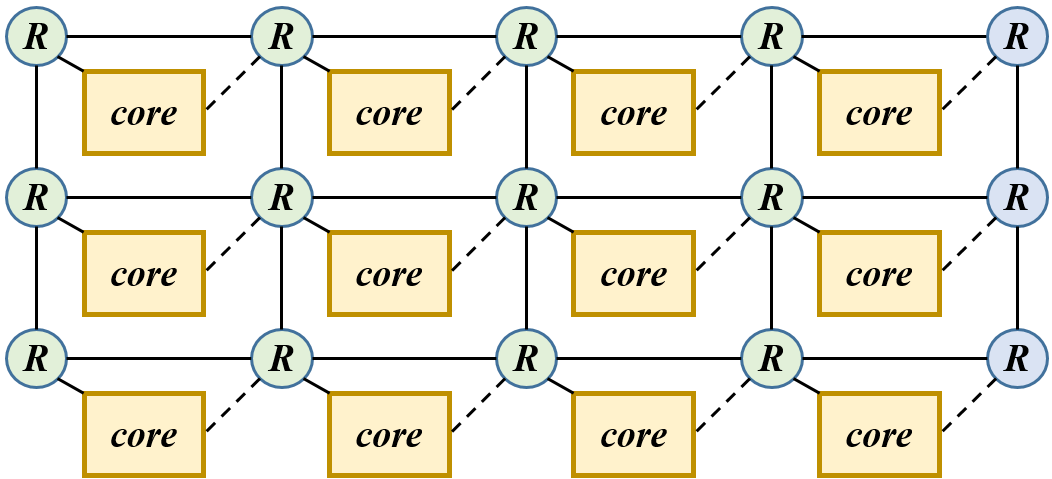}
        \caption{Router redundancy.}
        \label{fig2:sub3}
    \end{subfigure}
    \begin{subfigure}[b]{0.45\linewidth}
        \centering
        \includegraphics[width=\linewidth]{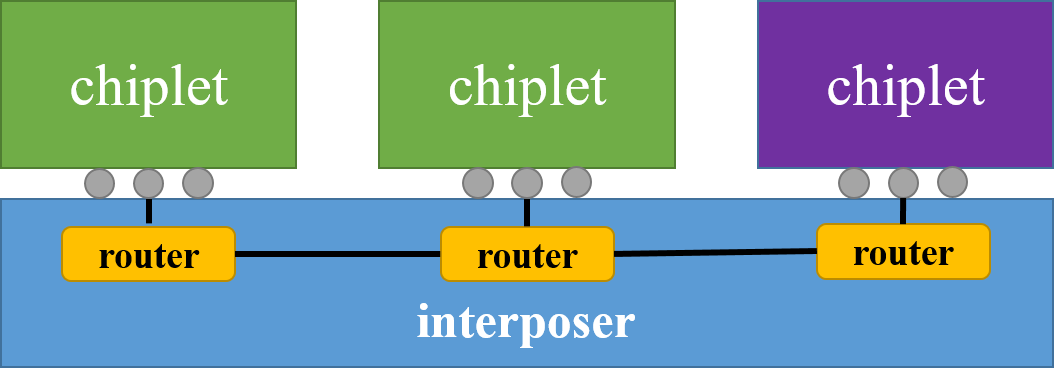}
        \caption{Chiplet redundancy.}
        \label{fig2:sub4}
    \end{subfigure}
    
    \caption{Illustration of chiplet-based architecture and redundancy assumptions. 
Each subfigure illustrates a model assumption: 
(a) single-chiplet integration in a 2.5D SCMS architecture; 
(b) module-level redundancy via additional core-router units; 
(c) router-level redundancy with $M$:$1$ rerouting; 
(d) inter-chiplet redundancy through spare chiplet bonding.
}
    \label{fig:fourplots}
\end{figure}

\begin{itemize} \item The chiplet-based architecture adopts a 2.5D single-chip multi-system (SCMS) architecture, where each chiplet integrates multiple cores interconnected via internal routing. A chip is considered functional if the number of interconnected working chiplet exceeds a predefined threshold.

\item The proposed framework incorporates two types of intra-chiplet redundancy strategies: module-level redundancy and routing-level redundancy. In our model, redundant components are assumed to possess identical characteristics to their original counterparts. Each chiplet consists of multiple modules, with each module comprising one core and one router. As illustrated in Fig.~\ref{fig2:sub2}, module-level redundancy is implemented by increasing the number of modules within a chiplet. In the presence of component failures, these redundant modules enhance the maximum number of interconnectable chiplets, thereby improving both manufacturing yield and architectural reliability.

Furthermore, we introduce an $M$:$1$ routing redundancy scheme for $M \times N$ routing networks, as depicted in Fig.~\ref{fig2:sub3}. In this configuration, each core is equipped with an alternate redundant routing line. Compared to cores, routers contribute significantly less to chiplet area and cost. However, a single router failure can sever a functional core’s connection to the rest of the network, rendering otherwise healthy compute resources unusable. This results in disproportionately high losses relative to the cost of the failed router. To mitigate this, routing-level redundancy allows cores to sequentially migrate to the next available router in the event of a routing failure within a row. This mechanism improves interconnect-level yield and enhances runtime fault tolerance within the chiplet network.

\item The framework also supports interchiplet redundancy, as illustrated in Fig.~\ref{fig2:sub4}. This strategy enhances the overall yield during the chiplet-to-interposer bonding stage, though it requires additional interposer area to accommodate redundant inter-chiplet interconnects.

The framework also supports inter-chiplet redundancy, as illustrated in Fig.~3(d), 
which improves overall package yield by providing spare chiplets during bonding, 
at the cost of a modest increase in interposer area.
In modern 2.5D integration schemes, industrial standards such as AIB 
\cite{AIB18} and UCIe \cite{UCIe22} already incorporate fine-granularity redundant 
bumps or lanes for interconnect repair. These mechanisms introduce very small silicon 
and cost overhead while substantially improving bonding yield and link reliability. 
Consequently, interconnect failures do not constitute the dominant lifetime-limiting 
factor in the lifecycle analysis considered in this work.

Therefore, our model focuses on chiplet-level redundancy, which is the primary 
architectural lever influencing system lifetime and LCE.
\end{itemize}
The design of our cost modeling framework is centered around the proposed \textit{ lifecycle cost effectiveness (LCE)} metric, which quantifies the amortized cost per unit of compute capacity over the operational lifetime of a chiplet-based architecture. 
A lower LCE indicates lower amortized cost per unit of delivered compute,
and therefore reflects better overall economic efficiency. Formally, LCE is defined as:
\begin{equation}
\text{LCE} = \frac{C_{\text{total}}}{\Phi_{\text{lifetime}}}
\end{equation}
where $C_{\text{total}}$ represents the total engineering cost, including both non-recurring engineering (NRE) and recurring engineering (RE) components. This component reflects fixed design and manufacturing overheads, as well as yield-sensitive variations in per-unit cost.  $\Phi_{\text{lifetime}}$ represents the cumulative compute throughput delivered over the chip’s operational lifespan, accounting for both sustained performance over time and the effective duration of operation. This quantity serves as the architecture’s lifecycle compute capacity (LCC), which is formally modeled in Section III-C. This denominator inherently captures the effects of reliability degradation, depreciation, and compute density on long-term performance delivery. By this definition, architectures with higher compute density or longer operational lifetimes yield greater $\Phi_{\text{lifetime}}$, thus lowering the LCE and improving cost-efficiency. Conversely, decreased reliability or inefficient redundancy strategies can shorten usable lifetime, lowering $\Phi_{\text{lifetime}}$, which leads to a higher LCE—indicating worse cost-efficiency.

The LCE metric is intentionally defined from a hardware-centric
perspective. It evaluates cost-effectiveness based on peak usable
compute capacity and lifetime behavior, rather than workload-
specific execution characteristics. This abstraction enables
architecture-level comparison of redundancy and chiplet
partitioning strategies independent of specific application
workloads.

As illustrated in Fig.~\ref{fig:overview}, the evaluation of LCE involves two primary modules in the framework:

\begin{enumerate}
    \item \textbf{Engineering Cost Modeling:} Estimates the NRE and RE costs of multi-chiplet architectures under different redundancy strategies.
    \item \textbf{Lifecycle Compute Capacity Modeling:} Models the architecture reliability, operational lifetime, and compute capacity over time. It incorporates the effects of redundancy-enhanced reliability and combines these with manufacturing cost results to evaluate long-term cost-effectiveness per unit of compute capacity.
\end{enumerate}

\subsection{Engineering Cost Modeling}

The proposed model analyzes the engineering cost of the chip in two stages. The first stage focuses on the chiplet-level analysis, where Monte Carlo simulations are used to evaluate the number of interconnected functional cores within each chiplet based on its design parameters. This allows the estimation of yield, cost, and reliability metrics at the chiplet level.

The second stage performs multi-chiplet architecture level analysis for multi-chiplet integration. In this stage, defective chiplets are filtered out through virtual yield screening, reducing the number of faulty chiplets entering the packaging phase and minimizing the loss of good chiplets. All chiplets that pass the test are called known good die (KGD). Monte Carlo simulations are then applied to assess the number of interconnected functional chiplets in the assembled system, enabling the calculation of overall yield, engineering cost, and other related metrics.
\begin{figure}[ht]
\centering
\fbox{%
\parbox{0.45\textwidth}{%
\textbf{Chiplet Yield Estimation Flow}
\begin{itemize}
    \item Randomly inject faults into components based on yield of each component.
    \item If a router fails, disconnect all its interconnect links.
    \item Search for the largest connected core cluster.
    \item If the maximum number of interconnections for good cores exceeds the required number of working cores for the chiplet, the chip is considered functional.
    \item Repeat the simulation to estimate the overall chip yield and related metrics.
\end{itemize}
}
}
\caption{The proposed chiplet yield estimation flow based on Monte Carlo simulation}
\label{fig:mc_yield}
\end{figure}

\begin{figure*}[!t]
    \centering
    \includegraphics[width=0.95\textwidth]{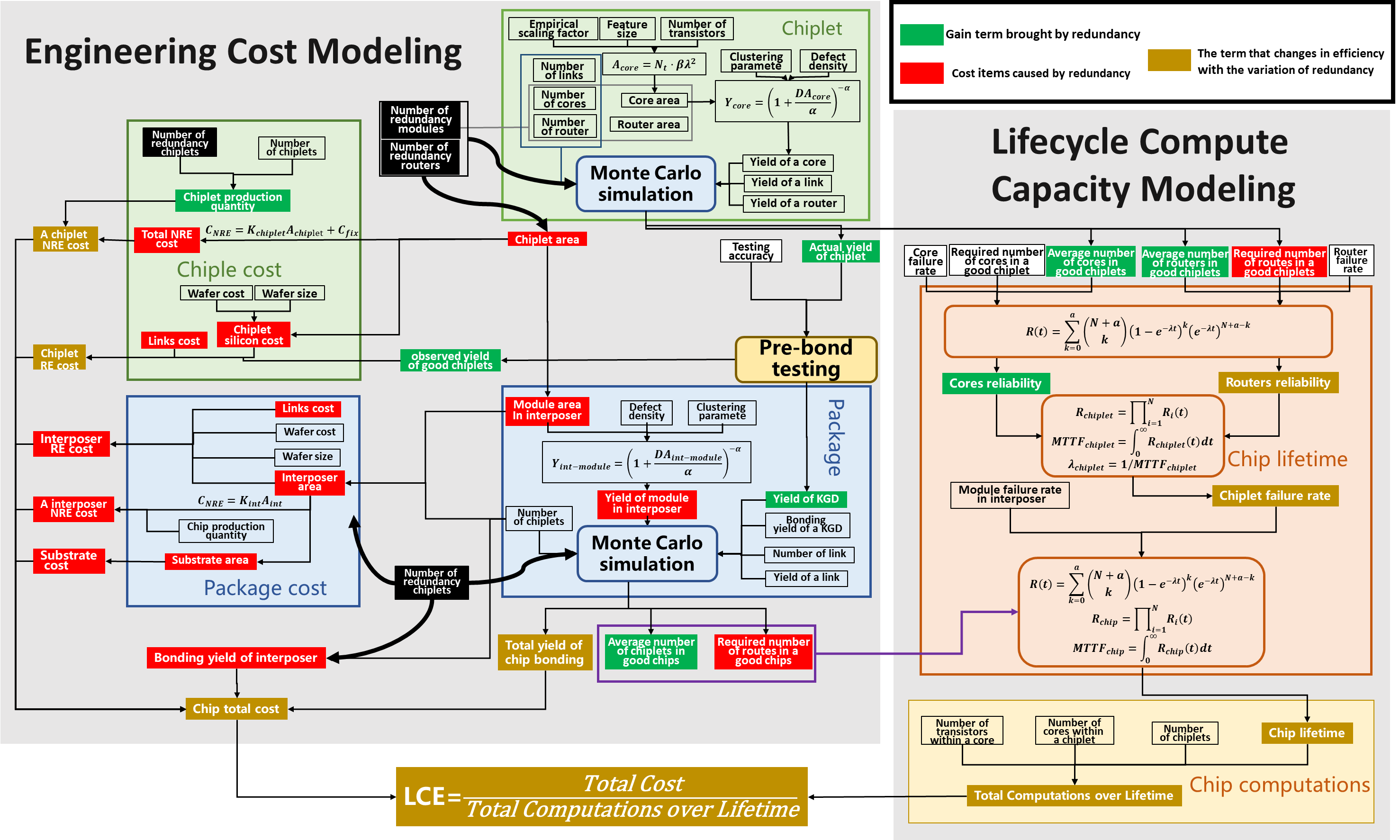}
    \captionsetup{skip=0pt}
    \caption{Overall Framework of the Cost-effectiveness Model. The framework integrates engineering cost and lifecycle compute capacity models to compute the LCE metric, with architecture configuration parameters (e.g., redundancy strategies) affecting yield, cost, and lifetime compute throughput.}
    \label{fig:overview}
\end{figure*}

\subsubsection{Chiplet}

We consider both the \textit{ recurring engineering (RE)} cost and the \textit{ non-recurring engineering (NRE)} cost for each individual chiplet.

The \textbf{RE cost} represents the necessary per-unit manufacturing expenses, including wafer fabrication and packaging for every chip produced. It can be modeled as:
\begin{equation}
C_{\mathrm{RE\text{-}chiplet}} = \frac{\sum C_{\mathrm{module\text{-}Si}} + C_{\mathrm{data}} \times N_{\mathrm{data}}}{Y_{\mathrm{ob}}}
\end{equation}
where $C_{\mathrm{module\text{-}Si}}$ denotes the silicon cost for manufacturing each module, $C_{\mathrm{data}}$ represents the cost of interconnecting data wires between modules, $N_{\mathrm{data}}$ is the number of data wires, and $Y_{\mathrm{ob}}$ is the observed chiplet yield.

Since the observed yield $Y_{\mathrm{ob}}$ significantly affects the amortized RE cost per functional chiplet, we model yield with greater granularity. Specifically, we decompose the chiplet into modules (e.g., cores and routers), and assume their yields are independent. The overall module yield is thus:
\begin{equation}
Y_{\mathrm{module}} = \prod_{\mathrm{components}} Y_{\mathrm{component}} = Y_{\mathrm{core}} \times Y_{\mathrm{router}}
\end{equation}

To accurately estimate the yield of each module, we adopt the \textit{Negative Binomial Distribution} (NBD) model. Unlike the Poisson model, the NBD captures defect clustering, providing more realistic yield estimations. For a core, the yield is given by:
\begin{equation}
Y_{\mathrm{core}} = \left(1 + \frac{D \times A_{\mathrm{core}}}{\alpha}\right)^{-\alpha}
\end{equation}
where $D$ is the defect density, $A_{\mathrm{core}}$ is the area of a core, and $\alpha$ is the clustering parameter. Both the defect density \(D\) and the core area \(A_{\mathrm{core}}\) are process-dependent. Core area can be further expressed as a function of transistor count and process technology:
\begin{equation}
A_{\mathrm{core}} = N_t \times \beta \lambda^2
\end{equation}
where $N_t$ is the number of transistors in the core, $\lambda$ is the feature size of the technology node, and $\beta$ is an empirical scaling factor that reflects layout efficiency. This formulation enables us to project yield from architectural parameters, such as logic complexity (via $N_t$) and fabrication process ($\lambda$).

However, for the entire chiplet, the overall chip yield $Y_{\mathrm{ob}}$ cannot be simply computed as the product of all module yields and link yields. This is because interconnectivity and structural redundancy play critical roles in determining whether the chiplet functions as intended. Therefore, we employ a \textbf{Monte Carlo simulation}-based method to estimate chiplet yield by performing network connectivity analysis under random fault injection. The specific procedure is shown in Fig.~\ref{fig:mc_yield}.


In each simulation iteration, component failures (cores and routers) are injected according to their individual yields. If a router fails, all of its connected links are disabled. As illustrated in Fig.~2-4, the faulty components are marked in red. After fault injection, we traverse the network to identify the largest set of interconnected functional cores. A chiplet is considered functional only if all cores are intact and fully connected through working routers.

By repeating this process over many iterations, we can estimate:
\begin{itemize}
    \item The chiplet manufacturing yield $Y_{\mathrm{ob}}$
    \item The average number of interconnected working cores per chip
    \item The average number of active routers
    \item The average number of essential routers required for full connectivity
\end{itemize}

To model redundancy, we incorporate both module-level and routing-level redundancy mechanisms into the simulation. For chiplets with redundant modules, the injected faults also apply to the redundant cores and routers, which are assumed to have identical characteristics to regular ones. After fault injection, the simulation counts the maximum number of interconnected, functional modules including redundant ones. If this number reaches the minimum required for normal operation, the chiplet is deemed functional.

For routing-level redundancy, we adopt an M:1 redundancy scheme, where one redundant router is appended per row in the routing network. Before checking core connectivity, router migration is performed: if a router fails, the associated core attempts to reroute through the redundant router in the same row. Since inter-row replacement is not allowed, if multiple routers fail within the same row, and no functional router remains, the corresponding core becomes disconnected.

The Monte Carlo simulation ultimately returns the estimated chiplet manufacturing yield, the average maximum number of interconnected cores, the average number of operational routers, and the average number of required routers for maintaining connectivity.

The Monte Carlo simulation is therefore extended to account for these redundancy-aware fault recovery rules. This allows us to estimate not only baseline yield, but also the \textit{effective yield under redundancy}, capturing the benefit of architectural fault tolerance in chiplet design.

The \textbf{NRE cost} of the chiplet consists of two components: one proportional to chip area, covering costs such as IP design, and the other independent of chip area, such as software licensing and equipment expenses. It can be expressed as:
\begin{equation}
C_{\mathrm{NRE-chiplet}} = K_{\mathrm{chip}}  \times A_{\mathrm{chip}} + C_{\mathrm{fix}}
\end{equation}
where $A_{\mathrm{chip}}$ is the chip area, $K_{\mathrm{chip}} $ is a coefficient determined by die area-dependent NRE costs, and $C_{\mathrm{fix}}$ represents area-independent NRE costs.

The NRE cost of the chip is amortized across all produced units. Under a fixed total NRE cost, the per-chip NRE cost is given by:
\begin{equation}
C_{\mathrm{NRE\text{-}chip}} = \frac{C_{\mathrm{NRE\text{-}total}}}{\mathrm{Volume}}
\end{equation}
where \textit{Volume} denotes the total production quantity of the chip.

\subsubsection{Engineering Model for Chiplet-Based Integration}

For chiplet integration technology, the manufacturing cost of an individual chiplet is comparable to that of a monolithic full chip. Chiplets are interconnected through micro-bumps and an interposer layer. The interposer is a thin silicon-based substrate that contains routers and physical links for inter-chiplet communication. It is connected to the package substrate via solder balls, which in turn connect the chiplet system to external circuits.

For 2.5D integrated chips, the RE cost and NRE cost can be modeled as:

\begin{equation}
C_{\text{RE}} = \frac{\sum C_{\text{RE-chiplet}} + C_{\text{int}} + C_{\text{sub}}}{ Y_{\text{chiplet-int}} \cdot Y_{\text{int-sub}}}
\end{equation}
\vspace{-8pt}
\begin{equation}
C_{\text{NRE}} = \sum C_{\mathrm{NRE-chiplet}} + C_{\mathrm{NRE-int}}
\end{equation}

Here, $C_{\text{int}}$ and $C_{\text{sub}}$ denote the cost of fabricating the interposer and the package substrate, respectively. The overall packaging yield is modeled as the product of two independent yield components: (1) the bonding yield between chiplets and the interposer, denoted as $Y_{\text{chiplet-int}}$, and (2) the bonding yield between the interposer and the package substrate, denoted as $Y_{\text{int-sub}}$.

We treat the inter-chiplet connections as an on-chip interconnection structure, where each chiplet is regarded as a core module. However, because chiplets need to be bonded onto the interposer via micro-bumps, the chiplet yield must incorporate both the KGD yield and the bonding yield.

To minimize waste of good chiplets, \textit{wafer-level testing} is employed to identify and select chiplets that meet performance and reliability requirements. Given the test accuracy $Y_{\text{test}}$, the observed yield of good chiplets (i.e., yield after testing) accounts for both correctly identified good dies and defective dies that are misclassified as good.

Thus, the observed manufacturing yield and the effective \textit{KGD yield} can be estimated as:
\begin{equation}
Y_{\text{ob}} = Y_{\text{chiplet}} \cdot Y_{\text{test}} + (1 - Y_{\text{chiplet}}) \cdot (1 - Y_{\text{test}})
\end{equation}
\vspace{-8pt}
\begin{equation}
Y_{\text{KGD}} = \frac{Y_{\text{chiplet}} \cdot Y_{\text{test}}}{Y_{\text{ob}}}
\end{equation}

To evaluate the value of chiplet-level redundancy, we modularize the interposer in our model. Specifically, the interposer is structured similarly to a chip and divided into several modules based on the positions where chiplets are bonded. Each module contains its own routers and logic units, and the routers across modules are interconnected via physical links.

We assume that the yield of each interposer module can be modeled by a Negative Binomial distribution, similar to chiplet modules. Since the interposer contains only sparse active elements and links, and is fabricated using a more mature process, it tolerates more defects. Therefore, its defect density is assumed to be lower than that of chiplets under the same technology node.

During packaging, the final bonding yield between chiplets and the interposer is jointly estimated by a Monte Carlo simulation, considering chiplet bonding yield, Known Good Die (KGD) yield, interposer module yield, and interposer link yield.

\subsection{Lifecycle Compute Capacity Modeling}

This section presents a formal model of lifecycle compute capacity, which quantifies the total computation a chip can deliver over its usable lifetime. The capacity depends on two key factors: the operational lifetime and the compute density, approximated by the number of transistors within functional cores. A longer chip lifetime enables more computations to be executed over the chip's lifespan, while a greater number of transistors within the cores increases the chip’s ability to process more computations per unit time. In this work, the Lifecycle Compute Capacity (LCC) is defined as the product of the architecture’s expected operational lifetime (\( \text{MTTF}_{\text{arch}} \)) and the total transistor count of its active cores:
\begin{equation}
\text{LCC} = \text{MTTF}_{\text{arch}} \times \text{Total Transistor Count of Active Cores}
\end{equation}
This metric reflects the net computation capacity available throughout the chip’s functional life, by integrating two complementary effects:

\begin{itemize}
\item Lifetime improvements enabled by redundancy and reliability enhancement strategies, which extend operational duration;
\item Increased computational capability due to higher transistor counts and architectural scaling within the active cores. Here, computation capacity can be regarded as an upper-bound indicator of the architecture’s peak performance potential, since a higher transistor count in active cores generally enables higher theoretical throughput.  
\end{itemize}

Note that architectural costs and overheads are modeled separately and accounted for to avoid overestimating the practical benefits of redundancy and scaling.

The reliability modeling of multi-chiplet architectures builds upon that of individual chiplets. We first analyze the lifetime and fault characteristics of a single chiplet, which serves as the basis for modeling composite reliability in multi-chiplet systems that integrate multiple chiplets via interconnect structures.

\subsubsection{Reliability and Lifetime Modeling of a Single Chiplet}

To characterize chip lifetime, we focus on permanent, non-recoverable hardware
failures. The \textbf{Mean Time To Failure (MTTF)} serves as a key metric,
capturing the expected operational duration of the chip before failure. The
reliability of an individual component is commonly modeled using an exponential
distribution with failure rate~$\lambda$:
\begin{equation}
R(t) = e^{-\lambda t}
\end{equation}
Here, \( R(t) \) represents the probability that the component remains
operational at time \( t \).

For a monolithic chip consisting of several modules, the overall reliability
can be expressed as the product of the reliabilities of the individual modules.
If the total number of modules is \(N\), then:
\begin{equation}
R_\text{chip}(t) = \prod_{i=1}^{N} R_i(t)
\end{equation}

The \textbf{MTTF} of the architecture, representing the expected operational
time before failure, is given by the area under the reliability curve:
\begin{equation}
\text{MTTF} = \int_{0}^{\infty} R(t)\, dt
\end{equation}

In cases where a chip includes redundant modules, the system continues
to operate as long as the number of functioning modules exceeds the minimum
required for correct operation. If the chip includes \(a\) redundant modules
and requires at least \(N\) operational modules, reliability becomes:
\begin{equation}
R_\text{redundant}(t) =
\sum_{k=0}^{a} \binom{N+a}{k} (1-e^{-\lambda t})^{k} (e^{-\lambda t})^{N+a-k}
\end{equation}

This models the probability that no more than 
$a$
a out of 
$N+a$ modules fail within time 
$t$. This improvement is directly reflected in the extended MTTF, indicating a longer usable system lifetime under the same failure rate conditions.

While redundancy extends operational lifetime, it also introduces system-level overhead, such as increased resource usage, structural complexity, and potential performance penalties. These trade-offs are incorporated into the compute capacity model to avoid overestimating the practical benefits of redundancy.


\noindent \textbf{Throughput Degradation Under Progressive Unit Failures.}
The baseline LCC model adopts a fail fast assumption in which the chip becomes
non-operational once any functional unit fails. This assumption yields a clean
and analytically tractable formulation and provides a reasonable baseline for
lifetime analysis. However, many modern architectures can continue to deliver
useful compute service even when a subset of functional units has permanently
failed, with throughput degrading gradually rather than collapsing abruptly.

To capture this broader class of designs, we introduce an extended model that
allows the system to remain functional with degraded throughput as long as the
number of failed units does not exceed a failure-tolerant threshold. Consider a
chip with $N$ identical functional units, each with reliability
$R(t)=e^{-\lambda t}$ and failure probability $F(t)=1-R(t)$ at time $t$. Let the
system remain operational as long as no more than $K_{\max}$ units have failed.
The probability that the chip remains within this degradation-tolerant region is
\begin{equation}
R_{\text{deg}}(t)=
\sum_{k=0}^{K_{\max}}
\binom{N}{k} F(t)^k R(t)^{N-k},
\end{equation}
which represents the probability that the system is still capable of delivering
nonzero throughput at time~$t$.

To quantify the degree of throughput degradation, we define the \textit{degraded
throughput factor} $\eta(t)$ as the expected fraction of active functional units
out of the $N$ original units:
\begin{equation}
\eta(t)=
\frac{1}{N}\sum_{j=0}^{N}(N-j)
\binom{N}{j}F(t)^jR(t)^{N-j}.
\end{equation}
Here, $(N-j)$ represents the number of surviving units when $j$ units have
failed, and the binomial term gives the probability of this failure pattern at
time~$t$. The factor $\eta(t)$ therefore provides a smooth, time-dependent
measure of the fraction of usable compute resources.

Combining the probability that the system is still operational with the expected
fraction of surviving units yields the effective throughput contribution at each
moment in time. The resulting lifecycle compute capacity under degraded
operation is
\begin{equation}
\text{LCC}_{\text{deg}} = \text{MTTF}_{\text{deg}} \times \text{Total Transistor Count of Active Cores}
\end{equation}
\begin{equation}
\text{MTTF}_{\text{deg}}=
\int_{0}^{\infty}
\eta(t)\, R_{\text{deg}}(t)\, dt ,
\end{equation}
which integrates the available compute capacity over the architecture's usable
lifetime. This formulation offers an optional extension to the baseline model
and characterizes designs that continue to provide meaningful compute capability
even as progressive unit failures accumulate.

\subsubsection{Reliability and Lifetime Modeling in Multi-Chiplet Architectures}
For multi-chiplet architectures, reliability modeling must consider not only the individual chiplets but also the active components in the interposer or bridging layers that connect them. After determining the Mean Time To Failure (MTTF) of each chiplet, the corresponding failure rate \(\lambda\) can be derived as the reciprocal of the MTTF, i.e., \(\lambda = \frac{1}{\text{MTTF}}\). Each chiplet is treated as an independent component with its own failure characteristics.

The overall system reliability should consider not only each chiplet but also the interposer components that maintain functional interconnectivity between chiplets. These interposer modules are essential for communication and data transfer, and their failures can cause system-level breakdowns even if all chiplets remain operational. Therefore, the architecture is modeled as a set of independent components, including chiplets and interposer modules. Under the assumption of independence between chiplets and interposer components, the combined reliability of these components can be approximated by the product of their individual reliabilities:
\begin{equation}
R_{\text{component}}(t) = R_{\text{chiplets}}(t) \times R_{\text{interposer}}(t)
\end{equation}
This formulation provides an approximate reliability for the entire multi-chiplet system, considering both chiplet failures and interposer-level failures.

When redundancy is incorporated at the chiplet level (e.g., spare chiplets or functional partitioning), similar binomial reliability models apply, which can improve system reliability by tolerating a certain number of failures among chiplets. However, it is important to also consider the potential negative impact redundancy may have on the reliability of inter-chiplet interconnects due to increased complexity and resource contention.





Finally, lifecycle compute capacity is evaluated by jointly considering the impact of redundancy-enhanced lifetime and the compute resources available within functional chiplets over the architecture’s lifetime. Both beneficial effects of redundancy and its potential overhead on routing complexity are integrated into the capacity calculation, ensuring a comprehensive assessment of compute capacity at architecture design level over its operational lifespan.

\section{Model Validation and Chiplet Redundancy Exploration }
In this section, we first validate the correctness of our model against results reported in existing studies. Following this validation, we employ the model to analyze and compare the lifecycle cost-effectiveness (LCE) of different redundancy strategies at both intra-chiplet and inter-chiplet levels. Additionally, we investigate the impact of various chiplet partitioning schemes on the overall lifecycle compute effectiveness. Finally, a multi-objective Pareto optimization is conducted to explore the trade-offs under given compute capacity constraints.

All chip designs considered throughout this section are based on a 14 nm process technology, while other key parameters are adopted from prior works\cite{stow2016cost, feng2022chiplet, chang2011design }. For consistency, units of cost, lifetime, and compute cost-effectiveness are expressed in arbitrary units (a.u.).

\subsection{Model validation}
To validate the effectiveness of our proposed model, we compare its results with those from the Chiplet Actuary framework, which has been validated using AMD-reported data. As the original AMD dataset is not publicly available, the comparison is conducted using the open-source implementation of the Chiplet Actuary model \cite{feng2022chiplet}. The Chiplet Actuary framework represents a comparative model that follows a chip-last packaging flow in a 2.5D chiplet integration scheme. Since this framework only evaluates engineering costs and does not consider router or interconnect yields, both yields are set to 1 in our comparison. Other parameters, such as interposer cost and NRE cost, are also aligned to ensure a fair comparison.

\begin{figure}[htbp]
\centerline{\includegraphics[width=0.9\linewidth]{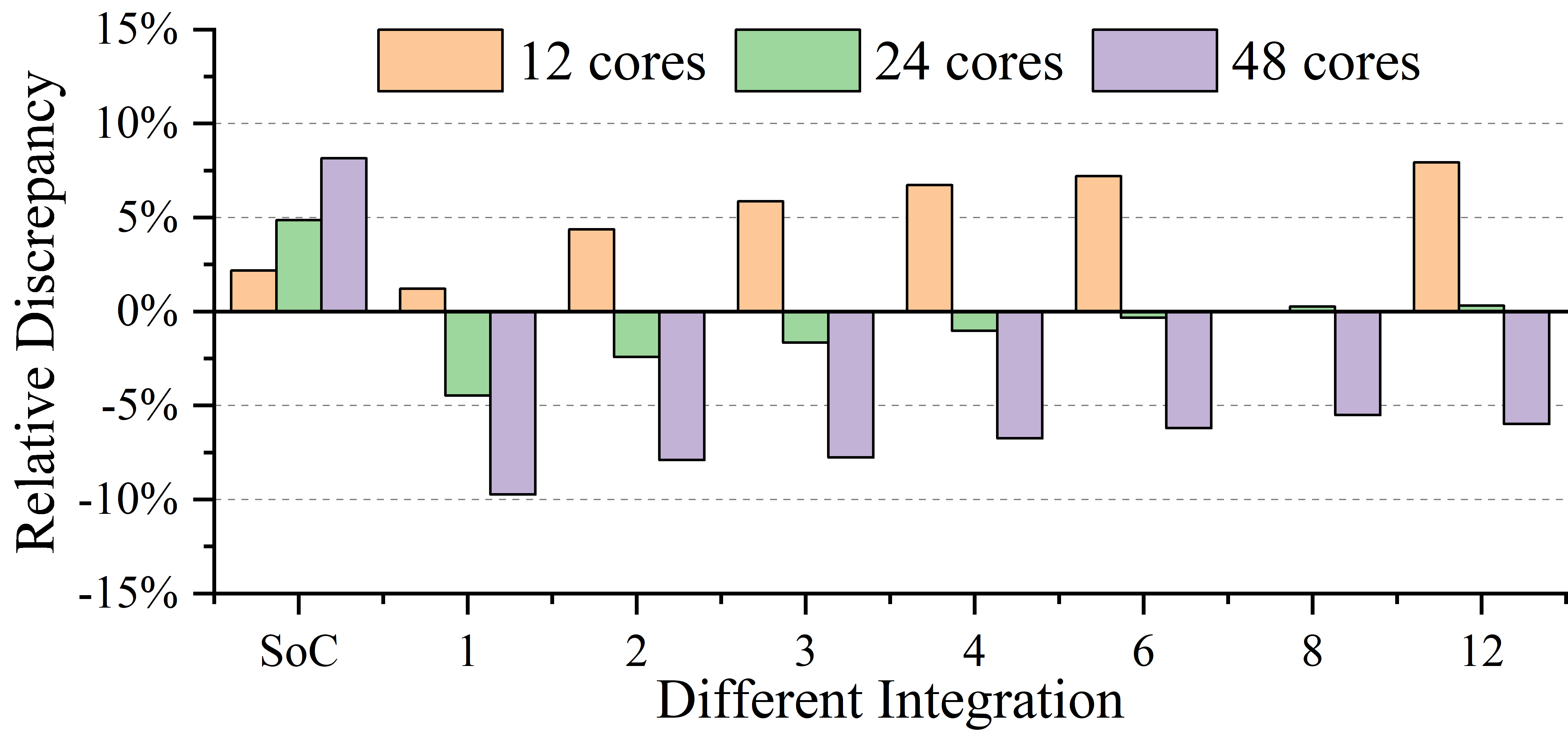}}
\caption{Comparison of engineering cost results between our model and the Chiplet Actuary framework.}
\label{Discrepancy}
\end{figure}

As shown in Fig.~\ref{Discrepancy}, the maximum deviation between our model and the Chiplet Actuary results is within 10\%, with an average deviation of 4.73\%. While Chiplet Actuary focuses only on engineering costs, our model extends this by introducing Lifecycle Compute Efficiency (LCE), which incorporates additional factors like computation capability and lifetime. 

Despite the observed deviation, the two models exhibit consistent relative behavior and bounded discrepancies across all evaluated configurations. This comparison demonstrates that our LCE framework is built on a solid foundation and validates its effectiveness in capturing cost trends across diverse multi-chiplet configurations.

\subsection{Lifecycle Cost-Effectiveness(LCE) Comparison of Different Intra-Chiplet Redundancy Strategies}

We first evaluate how module redundancy and router redundancy influence the proposed LCE metric, and then examine the underlying factors shaping these outcomes, including manufacturing yield, chiplet cost, total chip cost, and operational lifetime.

\begin{figure}[htbp]
\centerline{\includegraphics[width=0.8\linewidth]{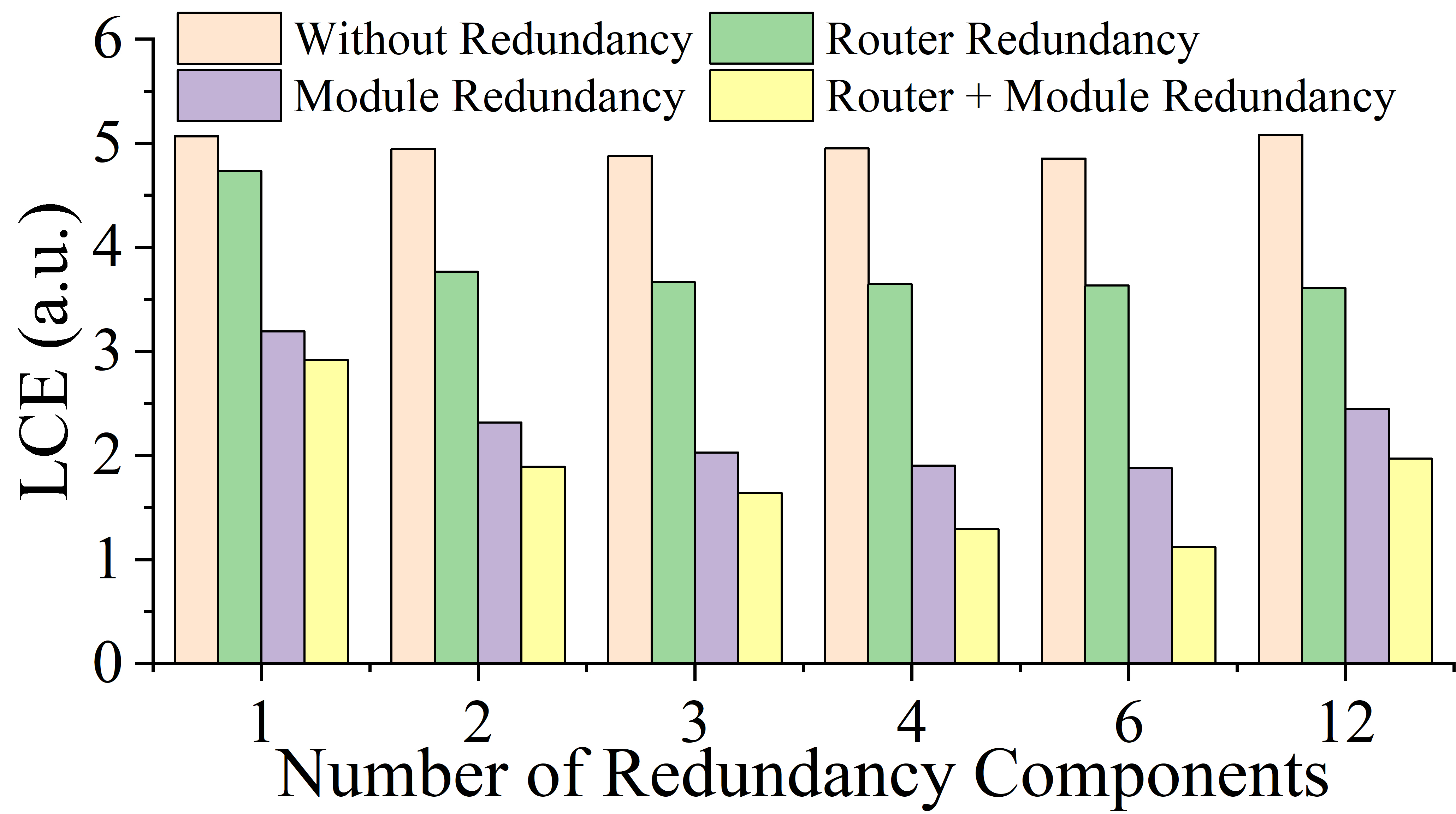}}
\captionsetup{skip=2pt}
 \caption{LCE Comparison of Different Intra-
Chiplet Redundancy Strategies (lower LCE indicates better cost-efficiency).}
 
\label{fig：intra_redundency}
\end{figure}

As illustrated in Fig~\ref{fig：intra_redundency}, for a 12-core chiplet, the lowest LCE is achieved when six redundant modules are deployed. In particular, even the introduction of a single redundant router incurs an optimistic economic benefit ratio. When applying module redundancy alone, the lowest LCE reaches 1.88, while router redundancy in isolation achieves a minimum LCE above 3.6. However, when both strategies are combined, LCE decreased significantly to 1.12. These results highlight that integrating both the module and router redundancy offers superior cost-efficiency and reliability compared to either approach used independently. 
While Fig~\ref{fig：intra_redundency} presents the aggregated Lifecycle Cost Effectiveness (LCE) across
different redundancy configurations, Fig.~\ref{fig:grid} and Fig.~\ref{fig:router_redundancy} decompose LCE into its
contributing factors, including yield, cost, and operational lifetime. This
decomposition is intended to explain the underlying causes of the observed
non-monotonic LCE trends rather than to replicate the LCE comparison itself.

\subsubsection{Module Redundancy: Yield-Driven Cost and Lifetime Trade-offs}

\begin{figure}[htbp]
    \centering
    \setlength{\abovecaptionskip}{5pt}
    \setlength{\belowcaptionskip}{0pt}

    \begin{subfigure}[b]{0.45\textwidth}
        \centering
        \begin{minipage}{0.49\textwidth}
            \centering
            \includegraphics[width=\textwidth]{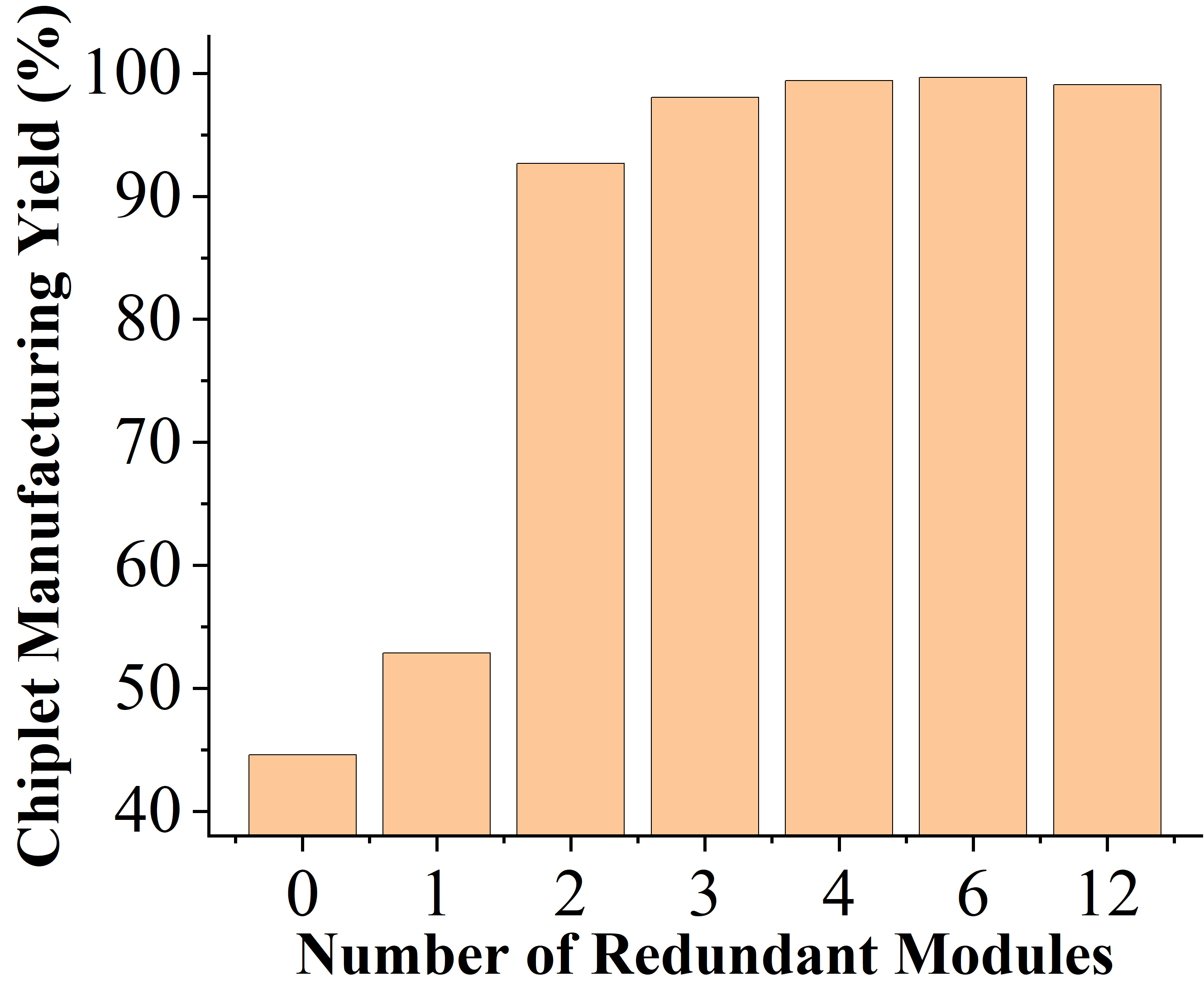}
        \end{minipage}
        \hfill
        \begin{minipage}{0.49\textwidth}
            \centering
            \includegraphics[width=\textwidth]{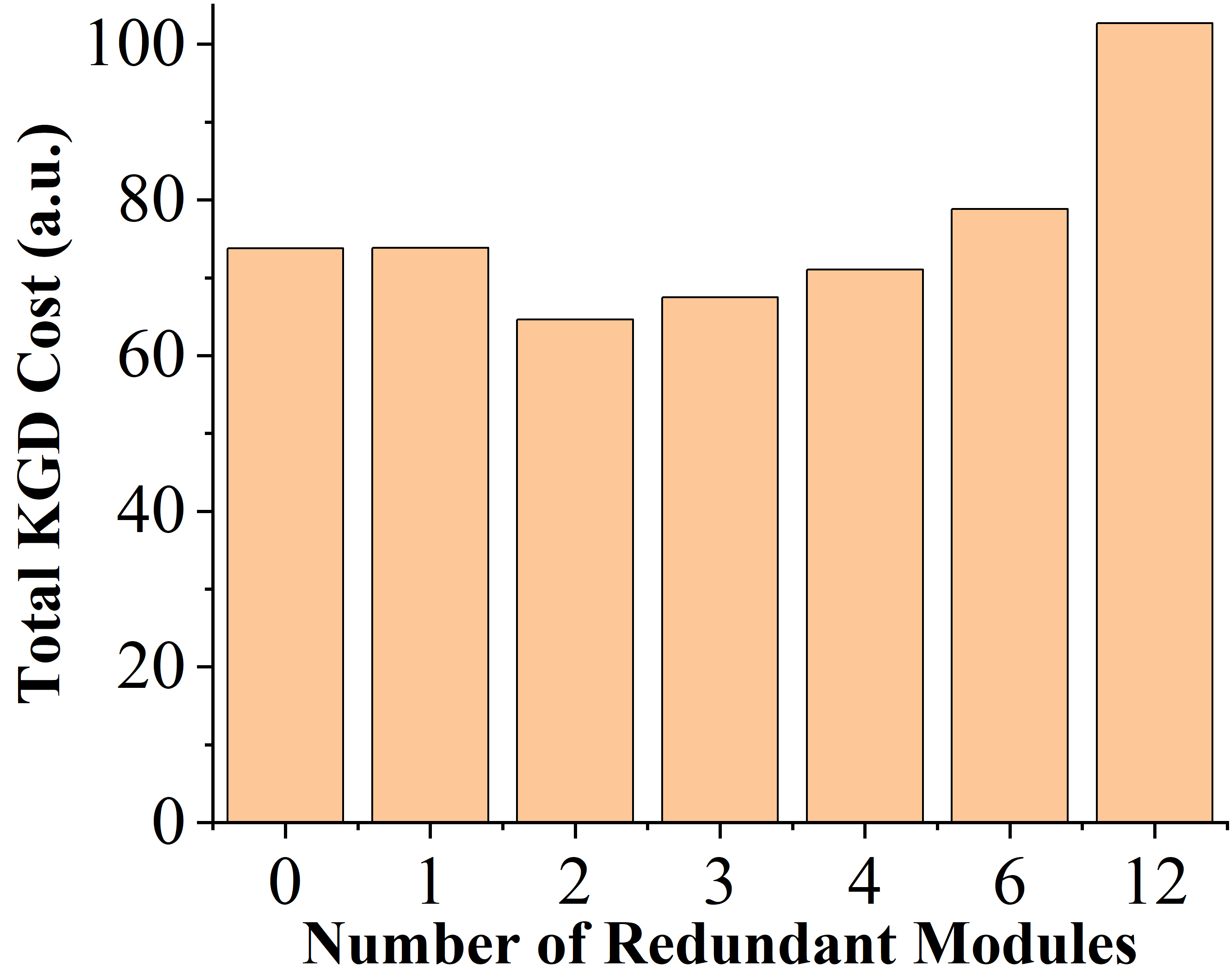}
        \end{minipage}
        \caption{Single Chiplet}
        \label{intra_redundency:sub1}
    \end{subfigure}
    \hfill
    \begin{subfigure}[b]{0.45\textwidth}
        \centering
        \begin{minipage}{0.49\textwidth}
            \centering
            \includegraphics[width=\textwidth]{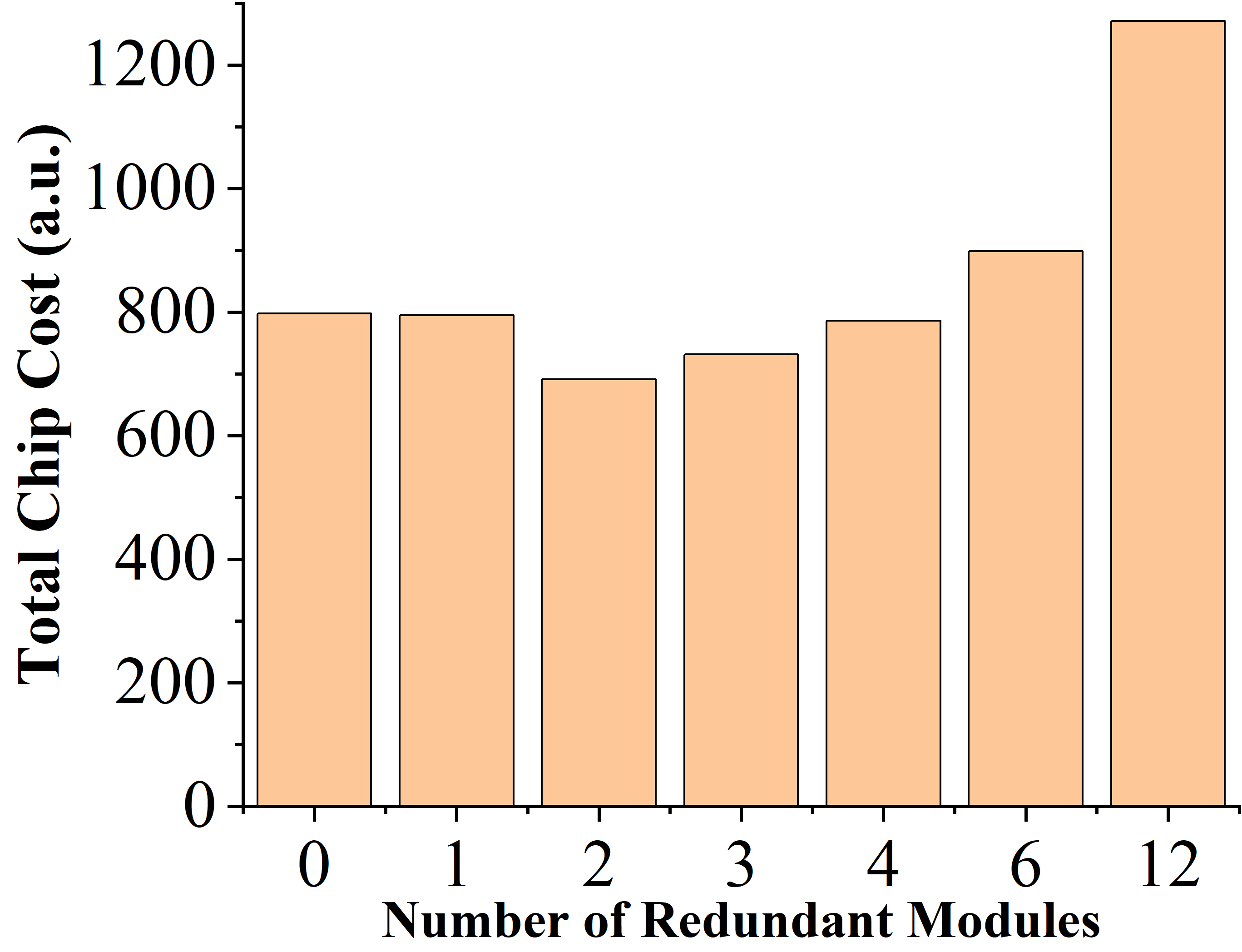}
        \end{minipage}
        \hfill
        \begin{minipage}{0.49\textwidth}
            \centering
            \includegraphics[width=\textwidth]{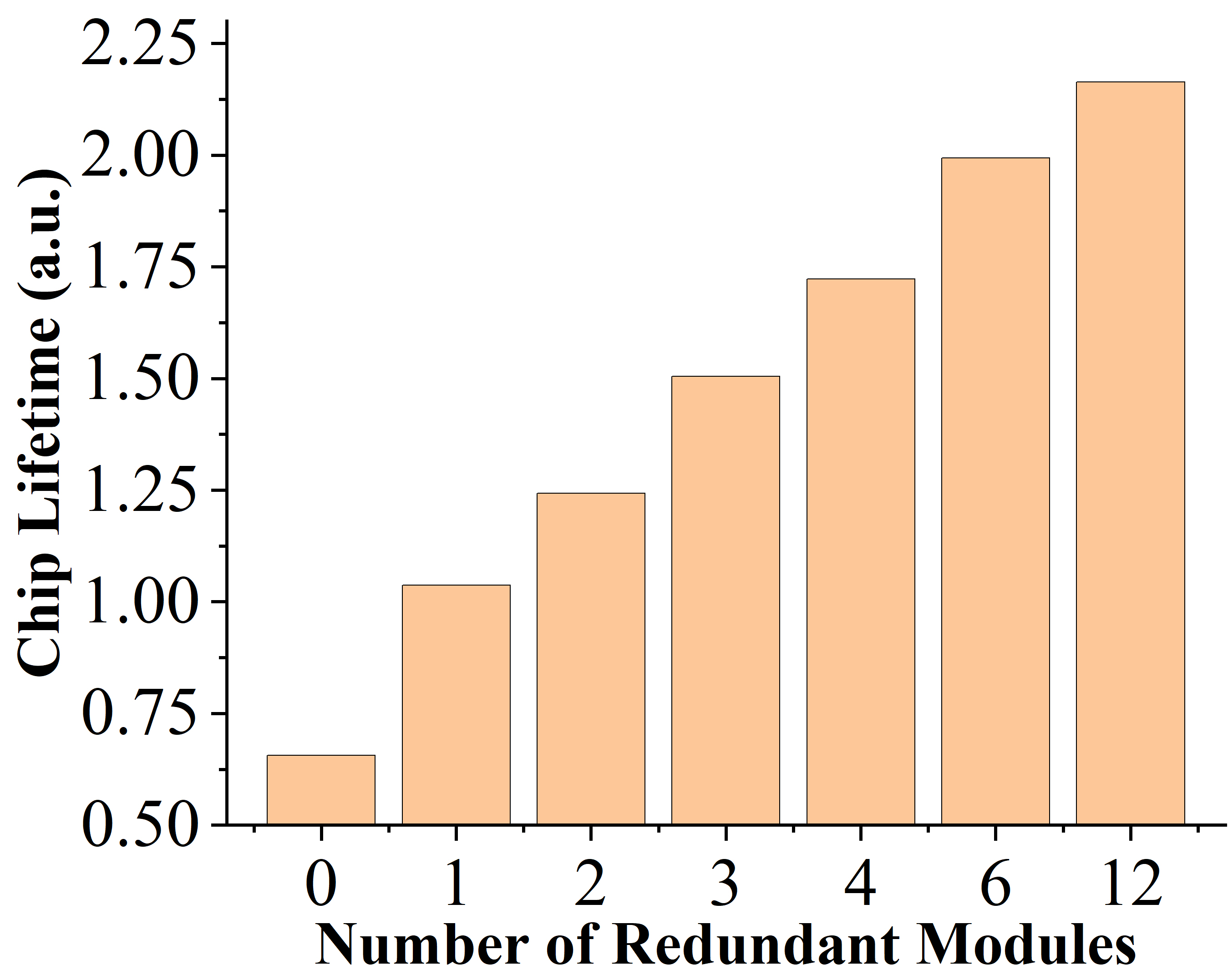}
        \end{minipage}
        \caption{Multi-chiplet Architecture}
        \label{intra_redundency:sub2}
    \end{subfigure}

    \caption{Impact of module redundancy on key metrics.}
    \label{fig:grid}
\end{figure}

To interpret these trends, we first analyze the impact of module redundancy at the chiplet and multi-chiplet architecture levels:

At the chiplet level, adding redundant modules improves manufacturing yield, though the marginal benefit decreases with each additional module. As shown in Fig.~\ref{intra_redundency:sub1}, redundancy initially reduces the cost per functional chiplet by boosting yield, but this advantage is gradually offset by increased area and manufacturing overhead. With six redundant modules, the chiplet cost ultimately surpasses the baseline.  This is because the chiplet yield is already very high, and the cost savings from further yield improvement through redundancy no longer compensate for the additional overhead introduced by the redundant modules.

At the multi-chiplet architecture level, the total chip cost is not simply the sum of the costs of individual chiplets. The extra packaging overhead caused by advanced packaging technologies and the reduced packaging yield significantly impact the total chip cost and cannot be ignored. Therefore, when discussing intra-chiplet redundancy, it is insufficient to consider chiplet cost alone; an analysis of the entire chip cost is necessary. 

As depicted in Fig.\ref{intra_redundency:sub2}, even though redundant modules introduce additional packaging overhead and lower packaging yield, they still provide economic benefits at the chip level. Redundant components can replace permanently failed components during chip operation, thereby extending the chip’s operational lifetime.  Fig.~\ref{intra_redundency:sub2} illustrates the impact of the number of redundant modules within a chiplet on the chip’s lifetime. As redundancy increases, the chip lifetime improves continuously, but the rate of improvement slows down with more redundancy. This slowdown is due to the increased complexity of the routing network required to ensure that redundant cores can properly substitute for failed ones. The chiplet’s reliability is constrained by the routing network’s reliability, which limits further improvements in chip lifetime. 

We observe that the trend of the chip’s LCE differs from the steadily increasing chip lifetime and does not align with chip cost changes. The LCE initially increases with the number of redundant modules, reaches a peak, and then starts to decline. Notably, the peak LCE does not coincide with the lowest chip cost, which occurs at two redundant modules. Although chip cost begins to rise beyond this point, the LCE does not decrease immediately because the lifetime gain grows faster than the cost increase. Under the influence of extended lifetime, the LCE reaches its maximum at six redundant modules. This indicates that extended operational longevity can more effectively amortize engineering costs.

\subsubsection{Router Redundancy: Enhancing Network Resilience and LCE}

\begin{figure}[htbp]
    \centering
    \setlength{\abovecaptionskip}{5pt}
    \setlength{\belowcaptionskip}{0pt}

    \begin{subfigure}[b]{0.45\textwidth}
        \centering
        \begin{minipage}{0.49\textwidth}
            \centering
            \includegraphics[width=\textwidth]{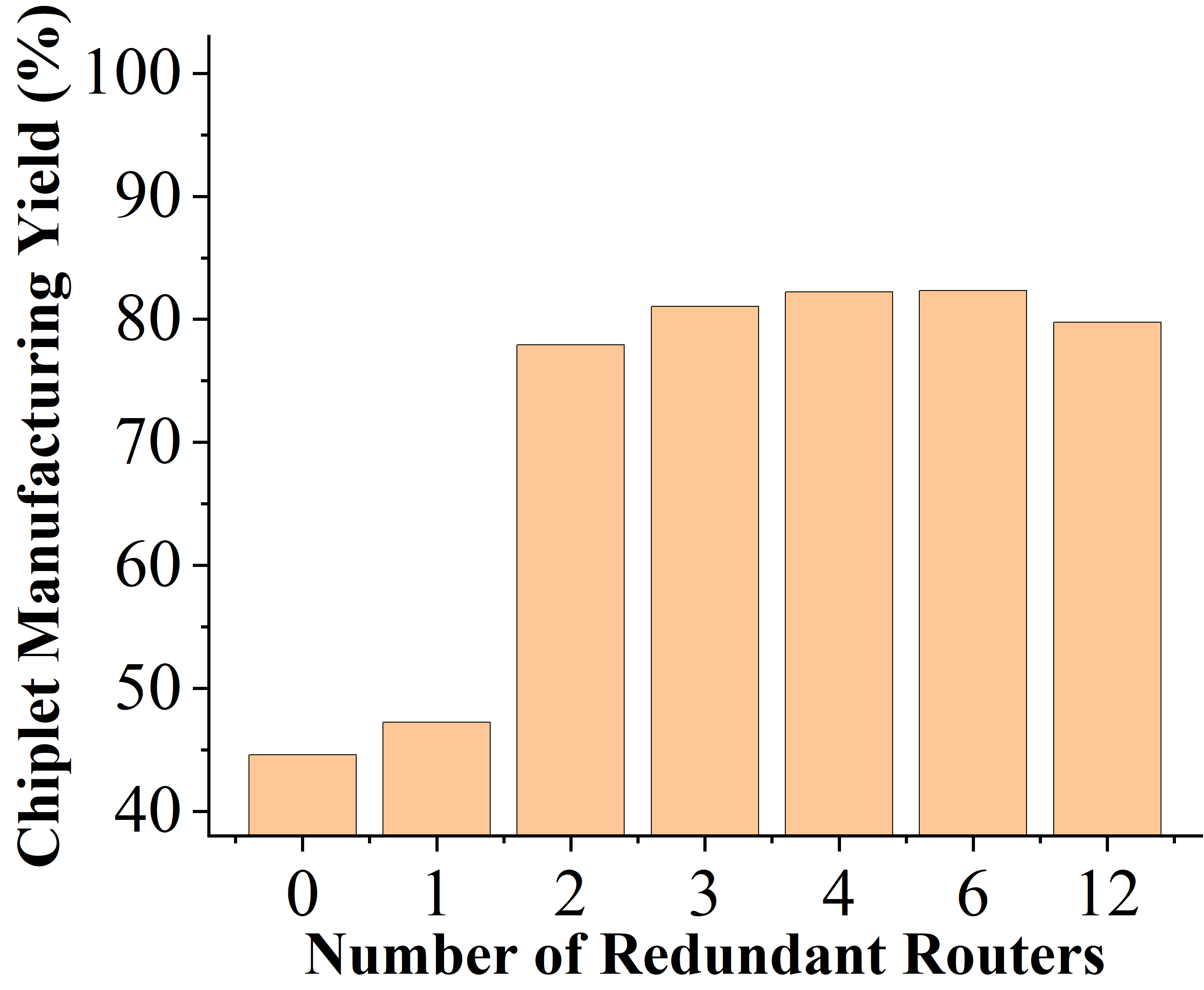 }
        \end{minipage}
        \hfill
        \begin{minipage}{0.49\textwidth}
            \centering
            \includegraphics[width=\textwidth]{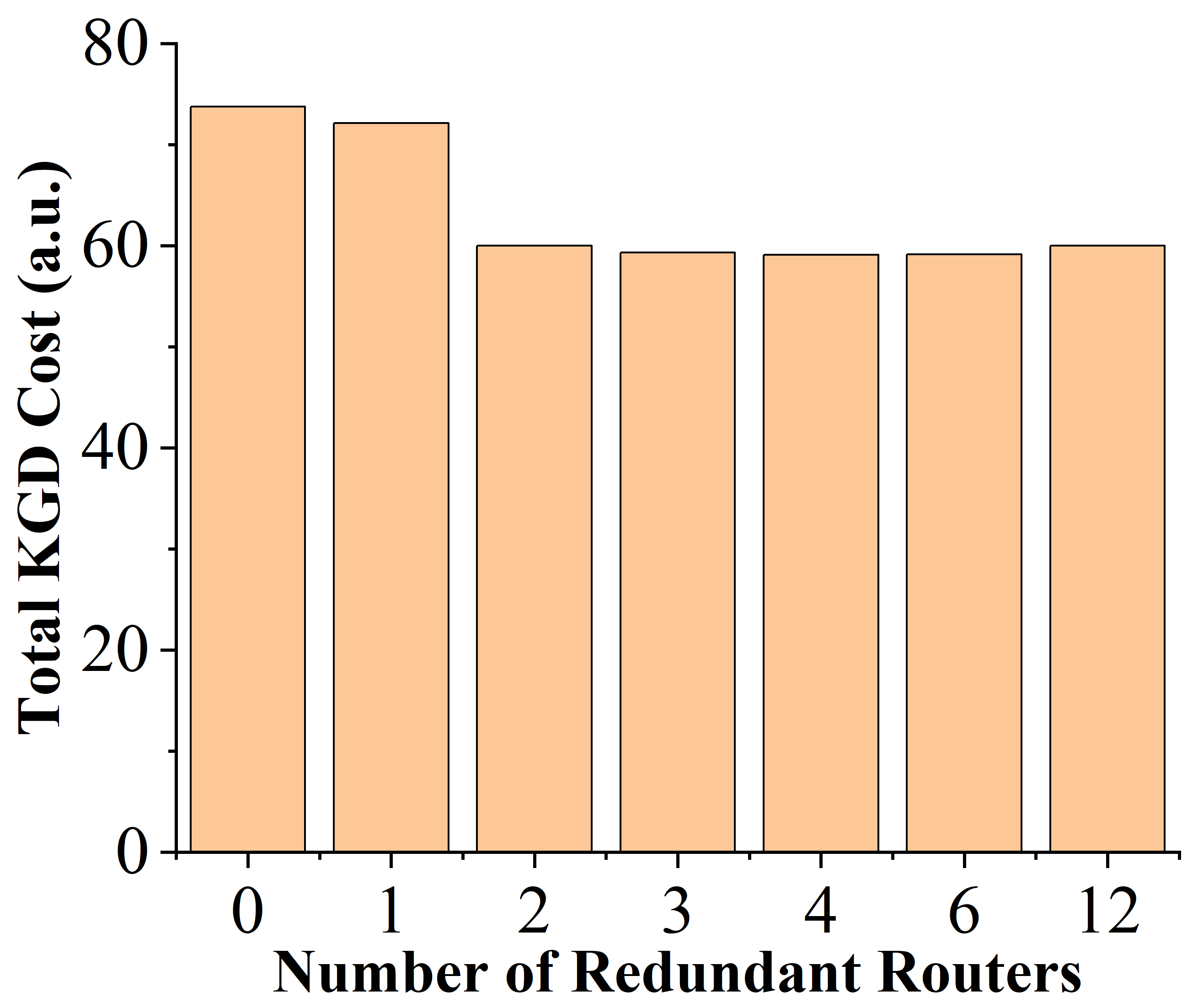}
        \end{minipage}
        \caption{Single Chiplet}
        \label{router_redundancy:sub1}
    \end{subfigure}
    \hfill
    \begin{subfigure}[b]{0.45\textwidth}
        \centering
        \begin{minipage}{0.49\textwidth}
            \centering
            \includegraphics[width=\textwidth]{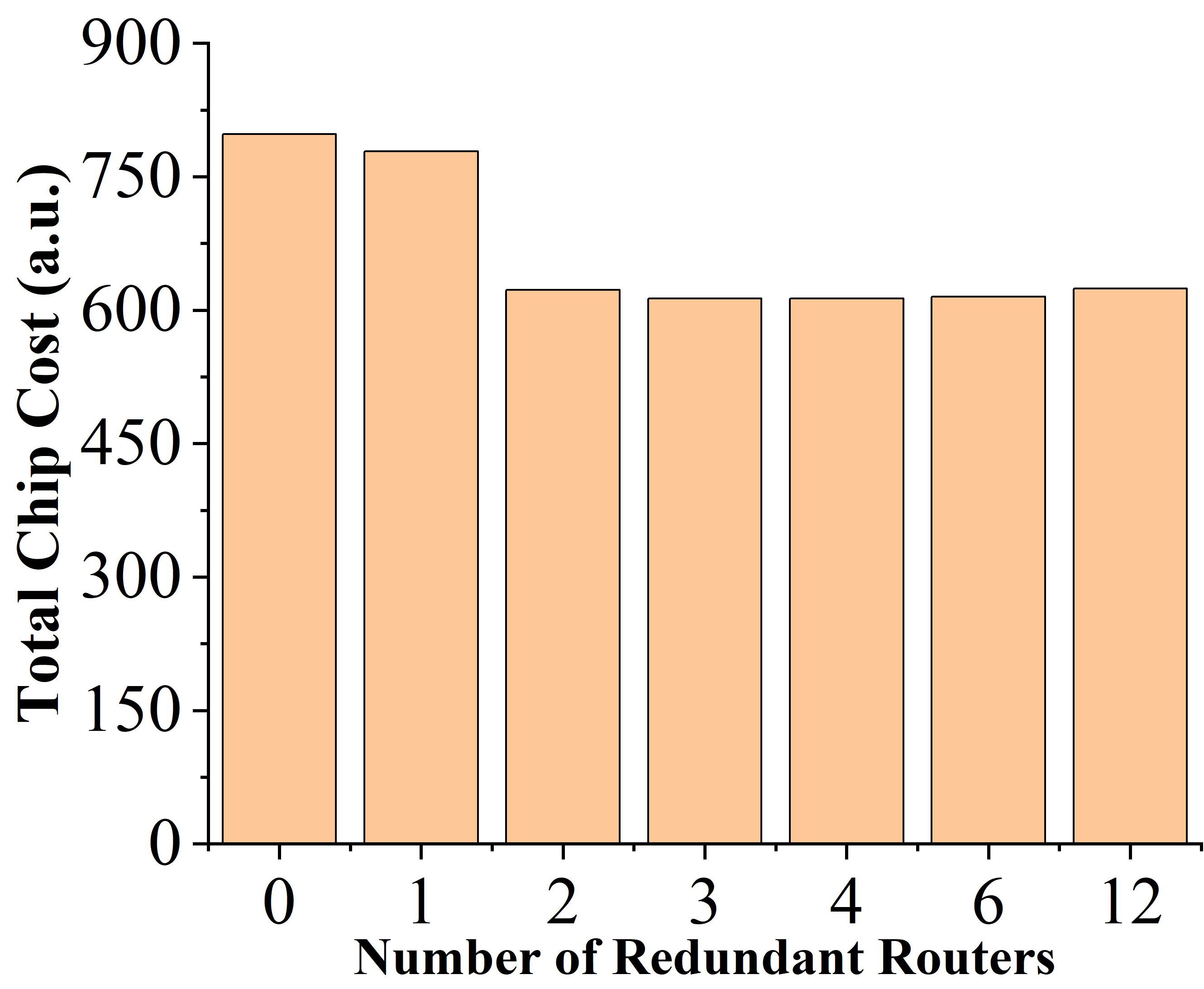}
        \end{minipage}
        \hfill
        \begin{minipage}{0.49\textwidth}
            \centering
            \includegraphics[width=\textwidth]{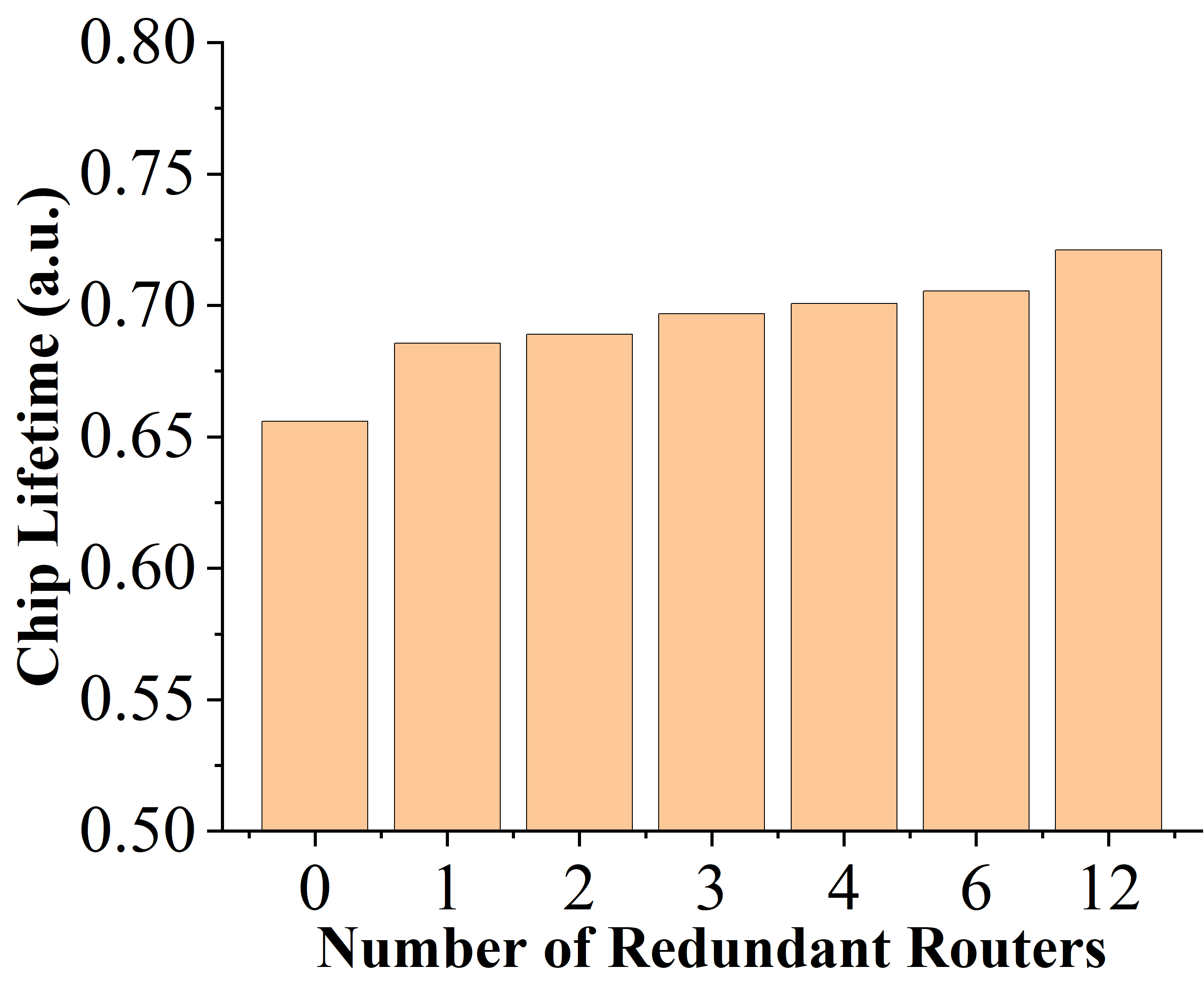}
        \end{minipage}
        \caption{Multi-chiplet Architecture}
        \label{router_redundancy:sub2}
    \end{subfigure}

    \caption{Impact of router redundancy on key metrics.}
    \label{fig:router_redundancy}
\end{figure}

We then analyze the impact of router redundancy, as shown in Fig.~\ref{router_redundancy:sub1}. At the chiplet level, introducing a moderate number of redundant routers improves manufacturing yield by mitigating the effects of individual router failures. However, excessive redundancy increases interconnect complexity and link failure probabilities, causing yield gains to plateau or slightly decline. Correspondingly, the cost per functional chiplet initially decreases as yield improves, reaching its lowest value with four redundant routers. Beyond this point, the area and routing overhead associated with additional routers outweigh the yield benefits, leading to increased chiplet cost.

At the multi-chiplet architecture level, router redundancy also affects operational lifetime and total chip cost. As shown in Fig.\ref{router_redundancy:sub2} illustrates that moderate router redundancy effectively reduces total chip cost by improving chiplet yield and increasing fault tolerance. However, as redundancy continues to increase, the additional cost and interconnect overhead gradually erode these benefits. Consequently, LCE reaches its optimum at a moderate redundancy level and degrades once the overhead surpasses the associated gains.

\subsubsection{Synergistic Effect: Combining Module and Router Redundancy}

While module redundancy improves manufacturing yield, it also increases the complexity of the network-on-chip (NoC), which can constrain chiplet reliability and limit the achievable lifetime benefits. Conversely, router redundancy enhances NoC resilience and extends lifetime without significantly impacting yield. This complementary nature motivates the combined use of both redundancy strategies to maximize chiplet performance and reliability.

\begin{figure}[htbp]
\centerline{\includegraphics[width=0.9\linewidth]{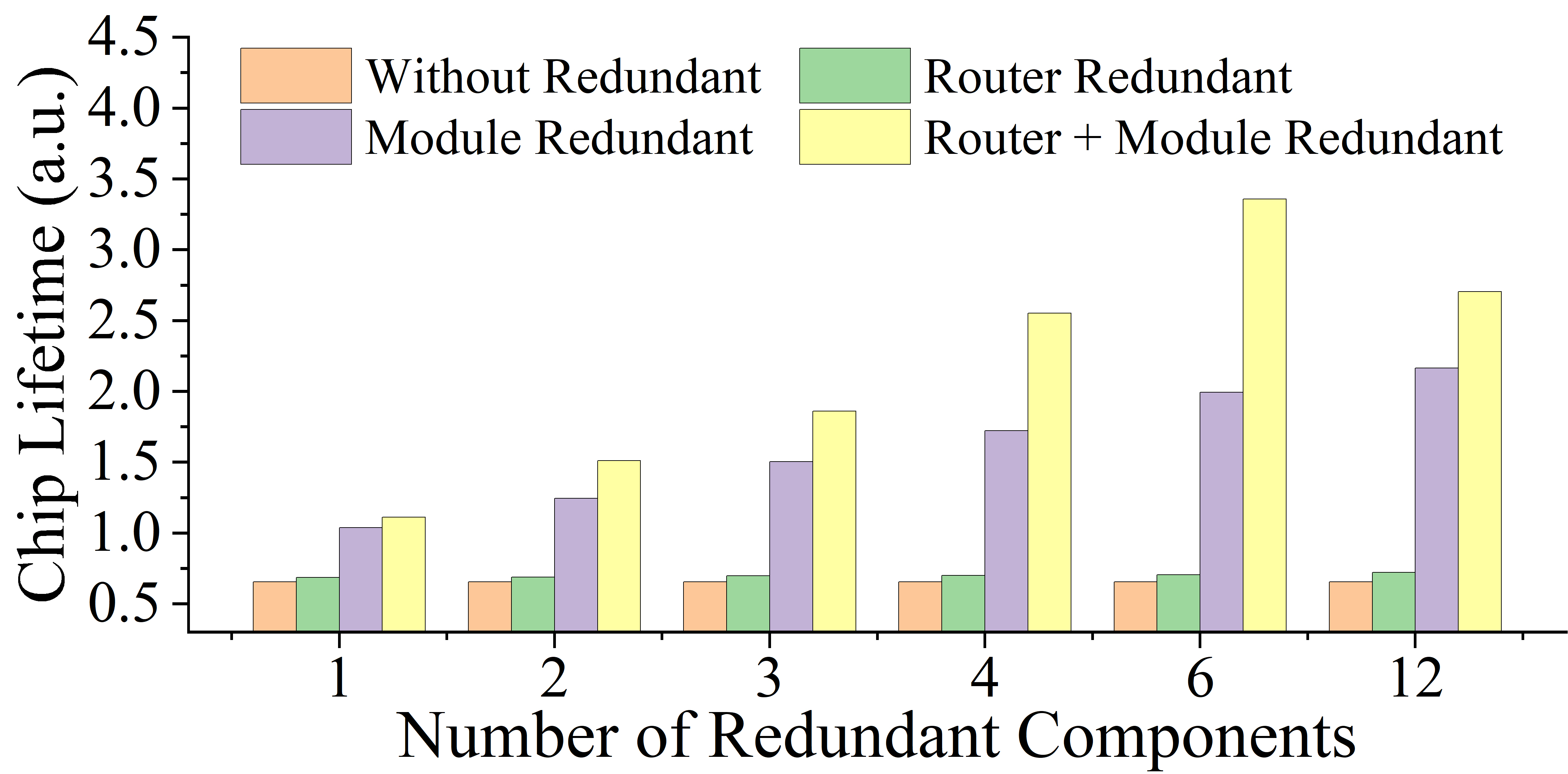}}
\caption{Lifetime analysis of the combined intra-chiplet redundancy strategies.}
\label{fig:lifetime}
\end{figure}

To verify this, we evaluated the chip lifetime when combining module and router redundancy, as shown in Fig.~\ref{fig:lifetime}. The results indicate that, regardless of the number of redundant modules, adding router redundancy consistently extends chip lifetime compared to cases without router redundancy. This demonstrates that the synergy between router and module redundancy yields significantly greater lifetime improvements than employing either strategy alone.

Building on these findings, we further assessed the impact of combining these redundancy strategies on the chip’s LCE. Consistent with Fig.~\ref{fig：intra_redundency}, the combined approach delivers more pronounced LCE improvements than single redundancy strategies, effectively balancing yield enhancement, NoC reliability, and runtime resilience. This integrated solution achieves superior cost-efficiency and operational longevity in multi-chiplet architectures.

\subsection{Lifecycle Cost-Effectiveness Comparison of Different Inter-Chiplet Redundancy Strategies.}
\begin{figure}[htbp]
\centerline{\includegraphics[width=0.9\linewidth]{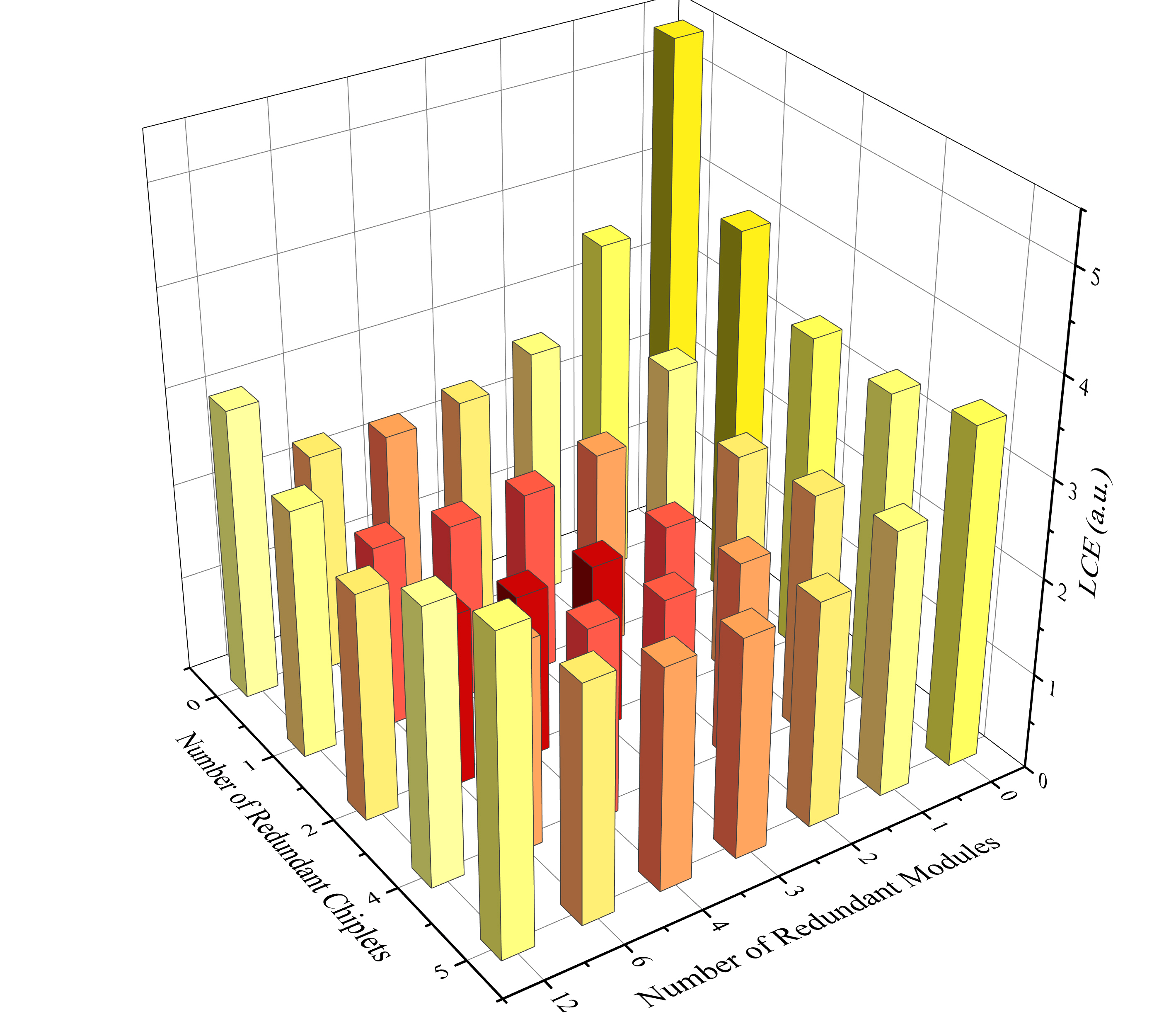}}
\caption{LCE Comparison of Different Inter-Chiplet Redundancy Strategies.}
\label{fig:Inter-Chiplet Redundancy}
\end{figure}

\begin{figure}[htbp]
\centerline{\includegraphics[width=0.9\linewidth]{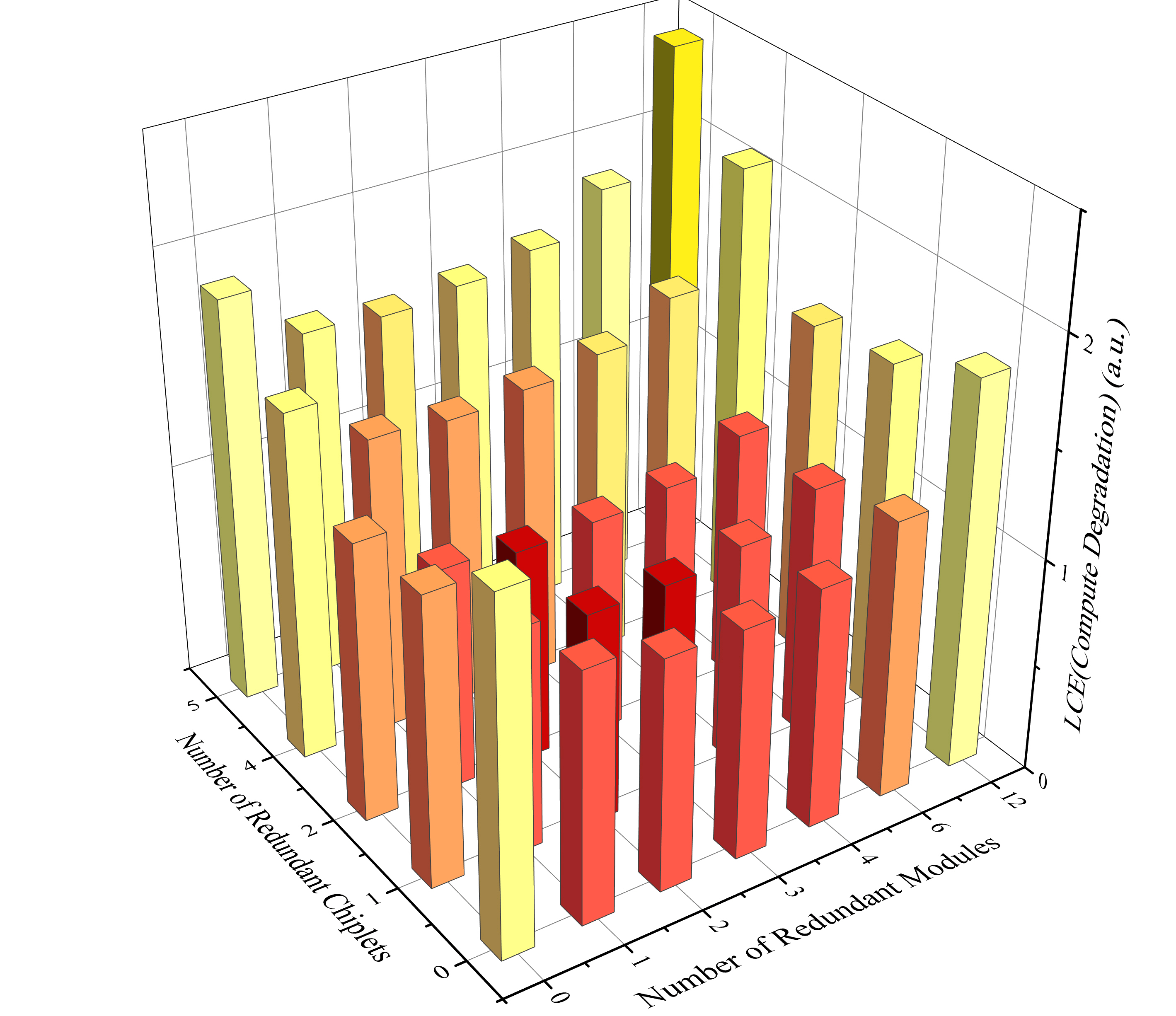}}
\caption{LCE(Compute Degradation) Comparison of Different Inter-Chiplet Redundancy Strategies.}
\label{fig:Inter-Chiplet Redundancy1}
\end{figure}

\begin{figure}[htbp]
    \centering
    \begin{subfigure}{0.9\linewidth}
        \centering
        \includegraphics[width=\linewidth]{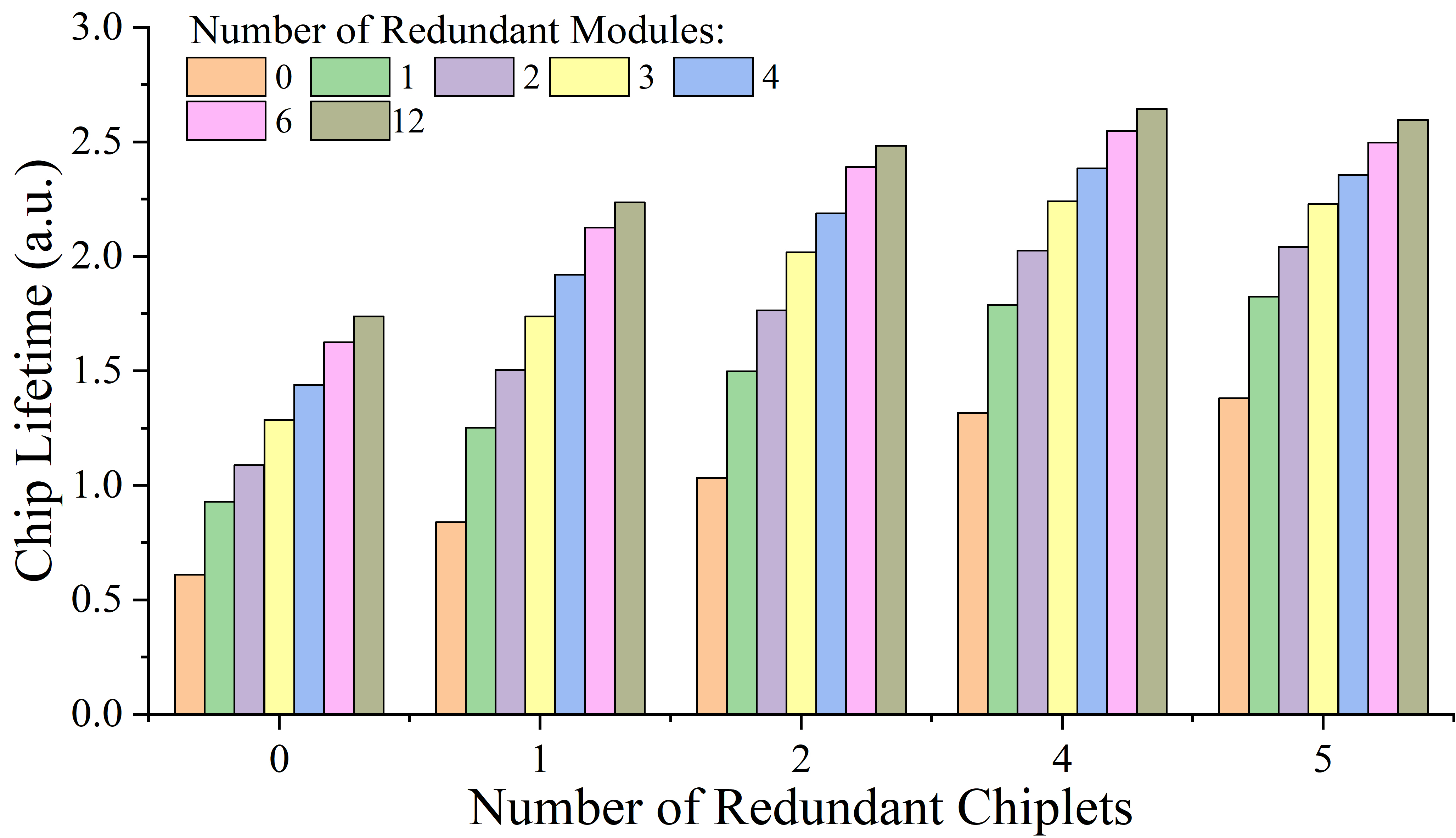}
        \caption{Chip lifetime variation with different redundancy strategies.}
        \label{fig:analysis_inter_re_cost}
    \end{subfigure}
    \hfill
    \begin{subfigure}{0.9\linewidth}
        \centering
        \includegraphics[width=\linewidth]{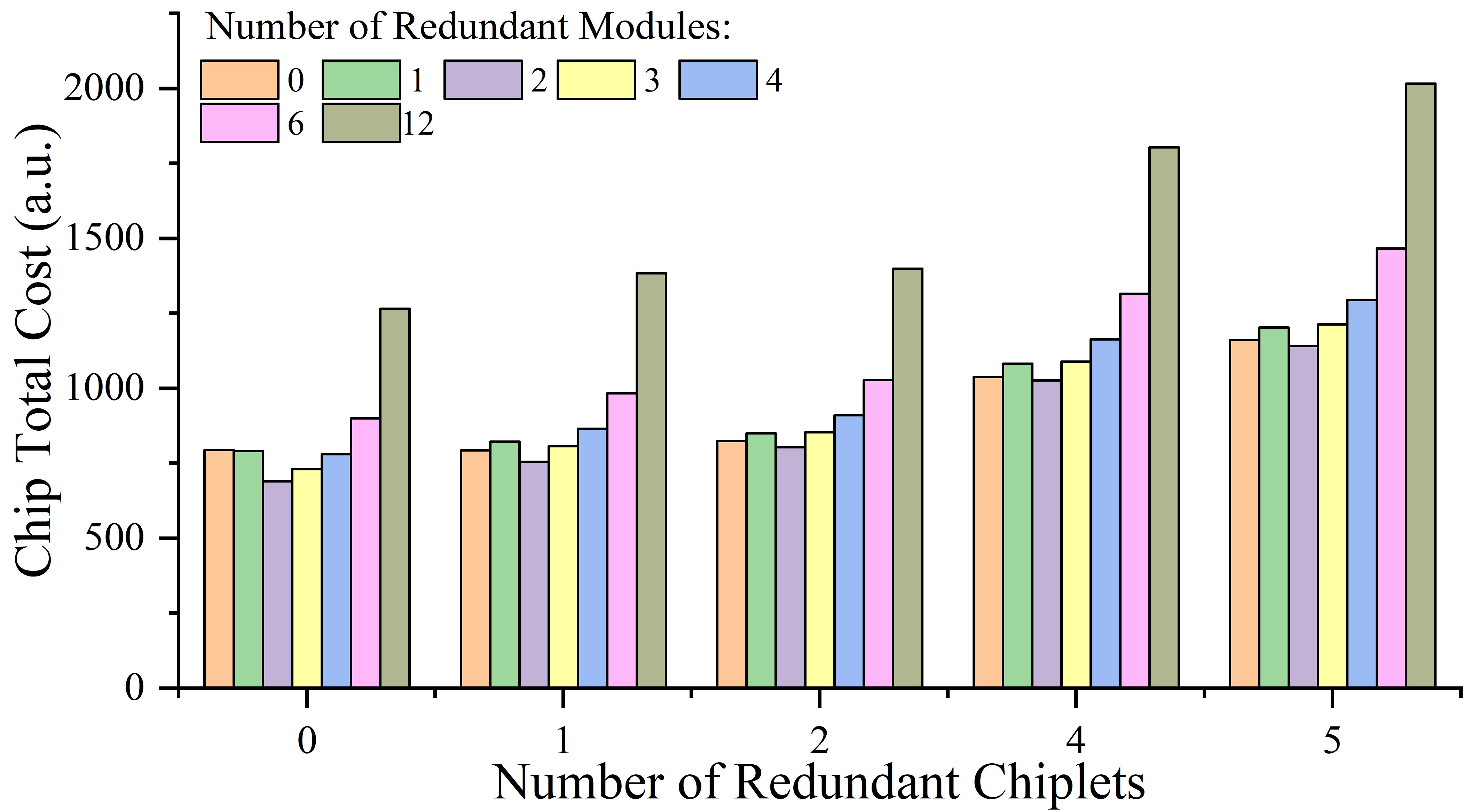}
        \caption{Chip total cost variation with different redundancy strategies.}
    \label{fig:analysis_inter_re_lt}
    \end{subfigure}
    \caption{Impact of Inter-Chiplet Redundancy on key metrics. }
    \label{fig:twosubfigures}
\end{figure}

To evaluate the cost-effectiveness of inter-chiplet redundancy, we performed extensive design-space exploration on a multi-chiplet architecture, varying the number of redundant chiplets and examining their interaction with intra-chiplet module redundancy. The 12-chiplet design serves as a hypothetical exploration point under a single-reticle assumption.

As shown in Fig.~\ref{fig:Inter-Chiplet Redundancy}, LCE first decreases with additional redundant chiplets, reaches a minimum, and then increases again. The lowest LCE occurs with four redundant chiplets, challenging the common assumption that inter-chiplet redundancy offers limited economic benefit.

To understand this behavior, we separately analyzed redundancy-induced changes in total engineering cost and operational lifetime. Fig.~\ref{fig:analysis_inter_re_cost} shows that adding redundant chiplets gradually increases manufacturing and integration costs due to additional packaging and interconnect resources. Meanwhile, Fig.~\ref{fig:analysis_inter_re_lt} shows that redundancy extends operational lifetime by enabling faulty chiplets to be replaced at runtime—an effect especially valuable when intra-chiplet NoC reliability and chiplet-level failures increasingly limit lifetime as more redundant modules are added per chiplet.

However, with a larger number of chiplets, the reliability of inter-chiplet communication becomes the dominant lifetime bottleneck. The expanded interconnect fabric introduces more failure points, reducing marginal lifetime gains and eventually causing LCE to worsen despite continued cost growth.

We further evaluated the combined effect of inter-chiplet and intra-chiplet redundancy. As shown in Fig.~\ref{fig:Inter-Chiplet Redundancy}, combining both strategies yields substantially lower LCE than using either alone, with the optimal configuration achieved at two redundant chiplets and four redundant modules per chiplet. This indicates that the two redundancy mechanisms interact and must be jointly optimized.

To complement the baseline evaluation, we further applied the degradation-aware lifecycle compute model from Section III, where the system continues to deliver partial throughput after module failures. The resulting LCE values (Fig.~\ref{fig:Inter-Chiplet Redundancy1}) are uniformly lower because the system accumulates additional compute output beyond the first failure. Nevertheless, the relative ranking of design points and key trends remain unchanged: the configuration with two redundant chiplets and four redundant modules per chiplet still provides the best lifecycle cost-effectiveness.

These results show that progressive degradation alters absolute LCE values but not the fundamental trade-offs. Across both fail-fast and degradation-aware models, joint optimization of inter-chiplet and intra-chiplet redundancy remains essential for maximizing cost-effectiveness in multi-chiplet systems.

\subsection{ Impact of Different Chiplet Partitioning Schemes on Lifecycle Compute Effectiveness}
\begin{figure}[htbp]
\centerline{\includegraphics[width=0.9\linewidth]{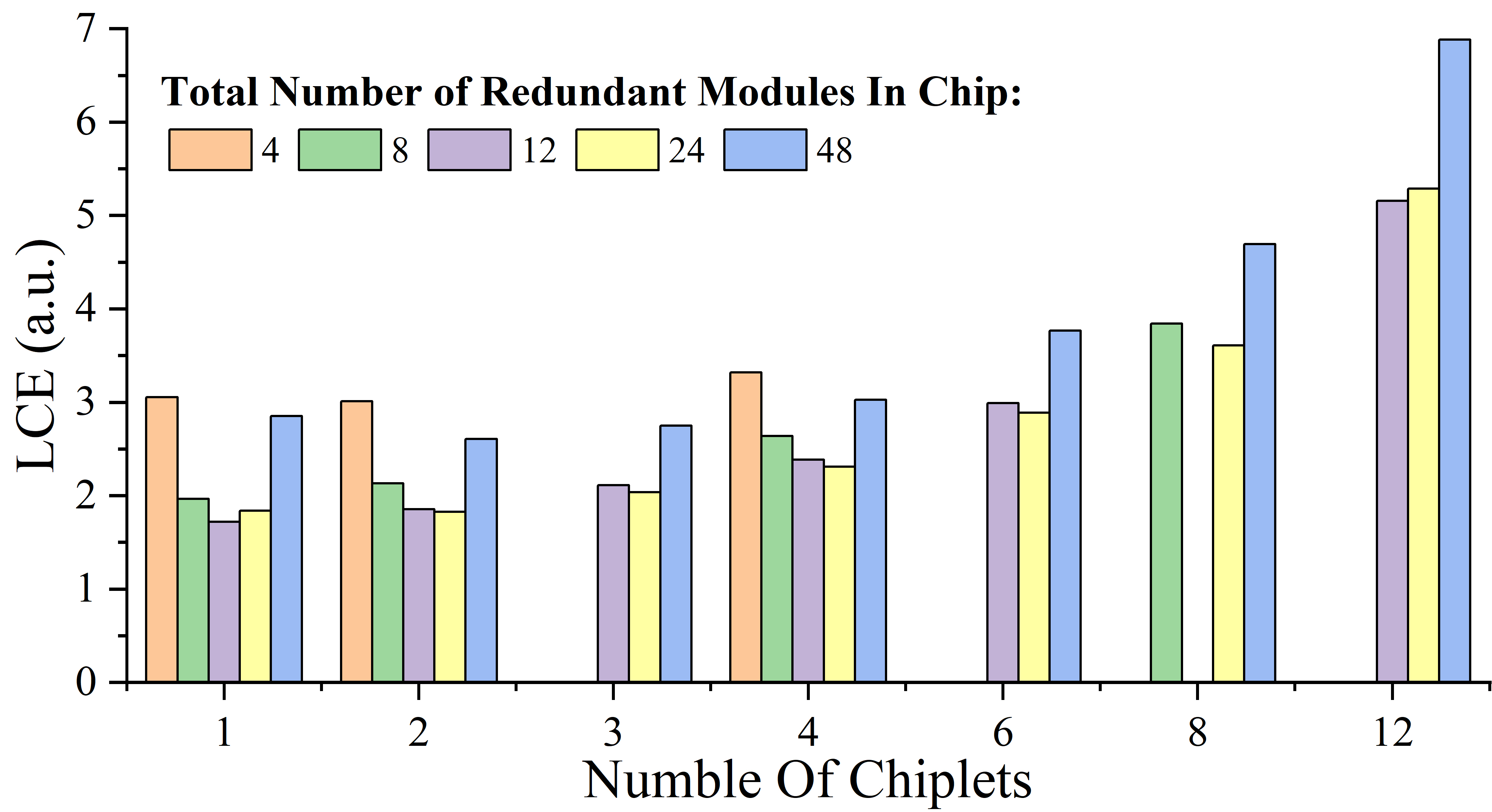}}
\caption{Impact of chiplet Partitioning schemes on LCE.}
\label{fig:integration}
\end{figure}

We further observe how different chiplet partitioning schemes affect the chip's lifecycle compute efficiency. To explore the most suitable partitioning scheme, we evaluated a 48-core chip with a fixed number of redundant modules under varying chiplet counts. 
Fig.~\ref{fig:integration}  indicates that chiplet partitioning and redundancy are tightly
coupled design dimensions. Different chiplet granularities exhibit different
LCE optima under different redundancy budgets, and for a fixed total number of
redundant modules, the choice of chiplet partitioning alone can substantially
change lifecycle cost-effectiveness. In particular, increasing the number of
chiplets generally reduces the effectiveness of redundancy by confining spare
resources within individual chiplets, while introducing additional packaging
and inter-chiplet reliability overheads. These results show that chiplet
partitioning and redundancy provisioning should be co-optimized, rather than
designed independently, to achieve optimal lifecycle cost-effectiveness.


Our results show that as the number of chiplets increases, the LCE of the architecture generally increases. This is because, in multi-chiplet architectures, redundant modules cannot repair permanent faults in other chiplets during runtime, reducing the effective utilization of redundancy resources and limiting their contribution to overall cihp lifetime extension.

When the number of redundant modules is set to 4, we observe that a two-chiplet configuration achieves a lower LCE than a monolithic single-chiplet architecture. In this scenario, the limited number of redundant modules is mostly consumed during manufacturing, leaving few available to handle runtime faults. Consequently, the redundancy’s contribution to operational lifetime is minimal, and the economic benefits of chiplet partitioning dominate, resulting in an minimum LCE at two chiplets.

As the chiplet count increases further, the economic advantage of smaller chiplets diminishes because of the reduced per-chiplet area and increasing packaging complexity. The utilization rate of redundant modules thus becomes the primary factor affecting LCE.

Interestingly, when the number of redundant modules increases to 24 or 48, this monotonic trend no longer holds. In these cases, the increased redundancy demands more complex intra-chiplet NoC routing networks to ensure effective fault tolerance, shifting the primary reliability bottleneck from core modules to the routing network. As a result, the lifetime benefits from additional redundant modules become limited, while the cost benefits of chiplet partitioning become more dominant, affecting the overall LCE trend.

\subsection{Multi-objective Pareto Optimization under Compute Capacity Constraints}

From the above analysis, LCE is a unified scalar metric that fundamentally captures the trade-off between total engineering cost and cumulative compute capacity delivered over the chip’s operational lifetime. This cost-effectiveness measure enables holistic optimization by balancing investment against peak compute potential and operational longevity. While abstracting many detailed factors, LCE guides optimization by reflecting design choices’ impact on achievable peak compute throughput under given cost constraints.

In this section, we investigate how optimal redundancy configurations adapt to varying compute capacity constraints, revealing implicit multi-objective optimization outcomes aligned with practical deployment scenarios.


\begin{figure}[htbp] \centerline{\includegraphics[width=0.9\linewidth]{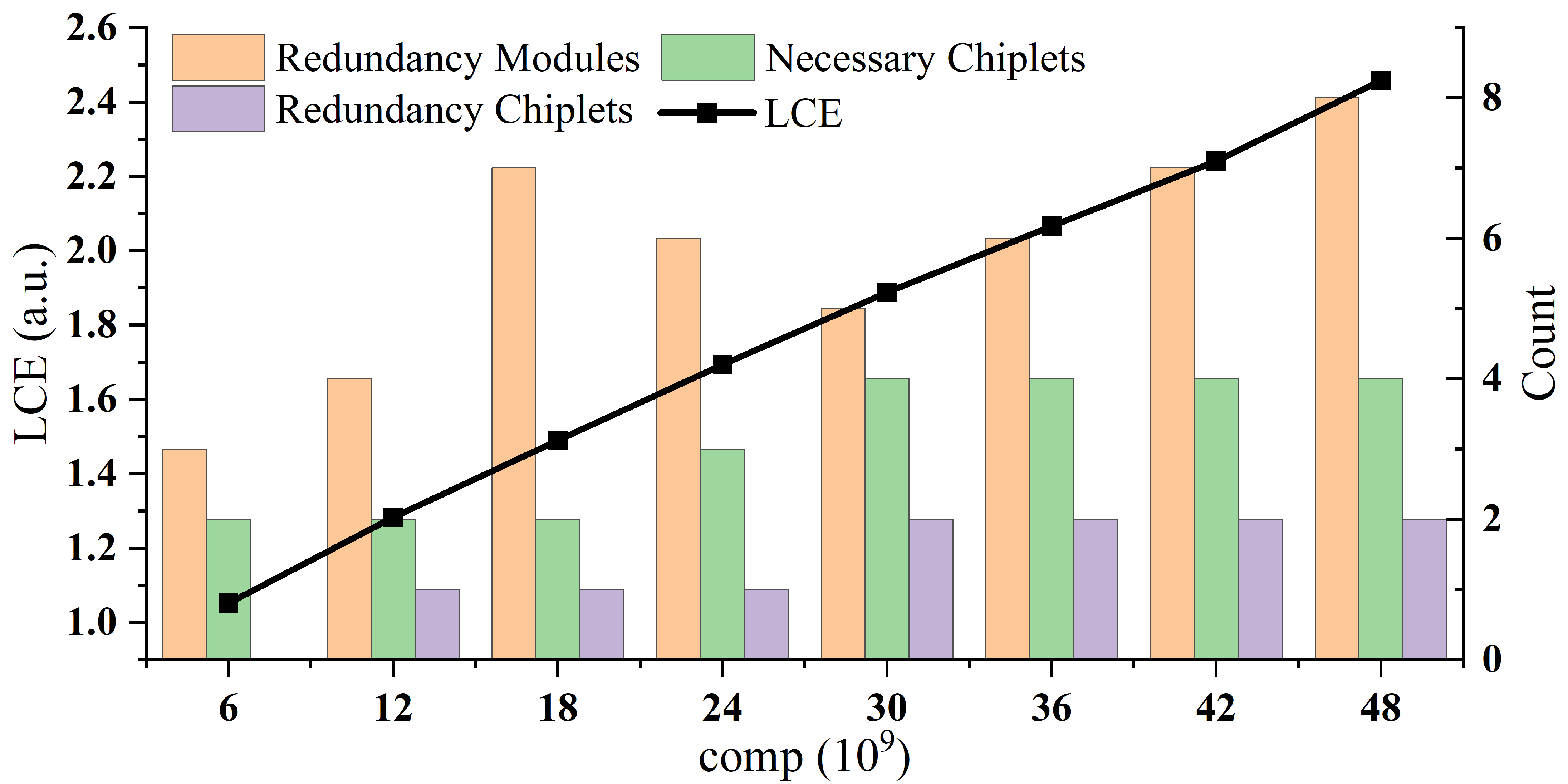}} \caption{Minimum LCE and corresponding parameters under different compute constraints.} \label{fig:optimization} \end{figure}

As shown in Fig.~\ref{fig:optimization}, the optimal redundancy configuration demonstrates two distinct scaling behaviors as compute demands increase. The number of active (2–4) and redundant (1–2) chiplets remains relatively stable, reflecting the diminishing marginal returns of chiplet-level redundancy. In contrast, module-level redundancy increases nonlinearly to address the elevated reliability challenges posed by larger chip designs.


This complementary adjustment forms a compute-aware Pareto frontier: increased module redundancy compensates for the reduced incremental benefits from additional chiplets. The combined effect leads to near-linear scaling of overall cost, as the fixed number of chiplets limits uncontrolled cost growth, while adaptive module redundancy targets reliability improvements. Consequently, this approach maintains the engineering principle that higher compute capacity requires proportionally greater investment, without incurring excessive overhead. The LCE metric converges to this optimal balance, reflecting the joint minimization of redundancy costs and maximization of lifetime benefits.

\section{Conclusion}
In this work, we have presented a comprehensive cost-effectiveness modeling framework tailored for multi-chiplet architectures, addressing a critical gap in evaluating the economic impact of redundancy strategies across multiple design layers. By introducing the Lifecycle Cost Effectiveness (LCE) metric, our approach goes beyond traditional cost models by jointly considering manufacturing expenses, yield improvements, and reliability-driven operational lifetime to provide an amortized view of compute cost.

Through extensive trade-off analyses and multi-objective optimization under varying compute capacity constraints, we demonstrated that optimal redundancy configurations require coordinated co-optimization at the module, router, and chiplet levels. This synergistic approach effectively balances yield, reliability, and economic efficiency, enabling the design of scalable multi-chiplet architectures without incurring prohibitive cost overheads, thereby validating the effectiveness of the proposed model in guiding economically optimized redundancy configurations. Our findings offer valuable insights for future chiplet-based system design, underscoring the importance of holistic, redundancy-aware cost modeling to navigate the complex trade-offs among manufacturing complexity, runtime reliability, and lifecycle economic efficiency.

\bibliographystyle{unsrt} 
\bibliography{reference}

@article{lu2024blending,
  title={Blending is all you need: Cheaper, better alternative to trillion-parameters llm},
  author={Lu, Xiaoding and Liu, Zongyi and Liusie, Adian and Raina, Vyas and Mudupalli, Vineet and Zhang, Yuwen and Beauchamp, William},
  journal={arXiv preprint arXiv:2401.02994},
  year={2024}
}

@inproceedings{naffziger2021pioneering,
  title={Pioneering chiplet technology and design for the amd epyc™ and ryzen™ processor families: Industrial product},
  author={Naffziger, Samuel and Beck, Noah and Burd, Thomas and Lepak, Kevin and Loh, Gabriel H and Subramony, Mahesh and White, Sean},
  booktitle={2021 ACM/IEEE 48th Annual International Symposium on Computer Architecture (ISCA)},
  pages={57--70},
  year={2021},
  organization={IEEE}
}

@inproceedings{wu2024advances,
  title={Advances and Reliability Challenges in Heterogeneous Integration in Chiplet ERA: From Solder to Copper to Optical Interconnects},
  author={Wu, Zhuo-Jie and Xu, Nan},
  booktitle={2024 Conference of Science and Technology for Integrated Circuits (CSTIC)},
  pages={1--4},
  year={2024},
  organization={IEEE}
}

@phdthesis{gehle2024reliability,
  title={Reliability and Resilience of Chiplets},
  author={Gehle, Erin Nicole},
  year={2024}
}

@inproceedings{ahmad2022heterogeneous,
  title={Heterogeneous integration of chiplets: Cost and yield tradeoff analysis},
  author={Ahmad, Mudasir and DeLaCruz, Javi and Ramamurthy, Anu},
  booktitle={2022 23rd International Conference on Thermal, Mechanical and Multi-Physics Simulation and Experiments in Microelectronics and Microsystems (EuroSimE)},
  pages={1--9},
  year={2022},
  organization={IEEE}
}

@article{li2020chiplet,
  title={Chiplet heterogeneous integration technology—Status and challenges},
  author={Li, Tao and Hou, Jie and Yan, Jinli and Liu, Rulin and Yang, Hui and Sun, Zhigang},
  journal={Electronics},
  volume={9},
  number={4},
  pages={670},
  year={2020},
  publisher={MDPI}
}

@article{shamshiri2011modeling,
  title={Modeling yield, cost, and quality of a spare-enhanced multicore chip},
  author={Shamshiri, Saeed and Cheng, Kwang-Ting},
  journal={IEEE Transactions on Computers},
  volume={60},
  number={9},
  pages={1246--1259},
  year={2011},
  publisher={IEEE}
}

@inproceedings{feng2022chiplet,
  title={Chiplet actuary: A quantitative cost model and multi-chiplet architecture exploration},
  author={Feng, Yinxiao and Ma, Kaisheng},
  booktitle={Proceedings of the 59th ACM/IEEE Design Automation Conference},
  pages={121--126},
  year={2022}
}

@inproceedings{stow2016cost,
  title={Cost analysis and cost-driven IP reuse methodology for SoC design based on 2.5 D/3D integration},
  author={Stow, Dylan and Akgun, Itir and Barnes, Russell and Gu, Peng and Xie, Yuan},
  booktitle={2016 IEEE/ACM International Conference on Computer-Aided Design (ICCAD)},
  pages={1--6},
  year={2016},
  organization={IEEE}
}

@inproceedings{stow2017cost,
  title={Cost-effective design of scalable high-performance systems using active and passive interposers},
  author={Stow, Dylan and Xie, Yuan and Siddiqua, Taniya and Loh, Gabriel H},
  booktitle={2017 IEEE/ACM International Conference on Computer-Aided Design (ICCAD)},
  pages={728--735},
  year={2017},
  organization={IEEE}
}

@inproceedings{jerger2014noc,
  title={NoC architectures for silicon interposer systems: Why pay for more wires when you can get them (from your interposer) for free?},
  author={Jerger, Natalie Enright and Kannan, Ajaykumar and Li, Zimo and Loh, Gabriel H},
  booktitle={2014 47th Annual IEEE/ACM International Symposium on Microarchitecture},
  pages={458--470},
  year={2014},
  organization={IEEE}
}

@inproceedings{chang2011design,
  title={On the design and analysis of fault tolerant NoC architecture using spare routers},
  author={Chang, Yung-Chang and Chiu, Ching-Te and Lin, Shih-Yin and Liu, Chung-Kai},
  booktitle={16th Asia and South Pacific Design Automation Conference (ASP-DAC 2011)},
  pages={431--436},
  year={2011},
  organization={IEEE}
}

@inproceedings{sapra2023exploring,
  title={Exploring multi-core systems with lifetime reliability and power consumption trade-offs},
  author={Sapra, Dolly and Pimentel, Andy D},
  booktitle={International Conference on Embedded Computer Systems},
  pages={72--87},
  year={2023},
  organization={Springer}
}

@inproceedings{wasala2023lifetime,
  title={Lifetime estimation for core-failure resilient multi-core processors},
  author={Wasala, Sudam M and Niknam, Sobhan and Pathania, Anuj and Grelck, Clemens and Pimentel, Andy D},
  booktitle={2023 IEEE 16th International Symposium on Embedded Multicore/Many-core Systems-on-Chip (MCSoC)},
  pages={293--300},
  year={2023},
  organization={IEEE}
}

@inproceedings{ehrett2021chopin,
  title={Chopin: Composing cost-effective custom chips with algorithmic chiplets},
  author={Ehrett, Pete and Austin, Todd and Bertacco, Valeria},
  booktitle={2021 IEEE 39th International Conference on Computer Design (ICCD)},
  pages={395--399},
  year={2021},
  organization={IEEE}
}

@inproceedings{taheri2022deft,
  title={DeFT: A deadlock-free and fault-tolerant routing algorithm for 2.5 D chiplet networks},
  author={Taheri, Ebadollah and Pasricha, Sudeep and Nikdast, Mahdi},
  booktitle={2022 Design, Automation \& Test in Europe Conference \& Exhibition (DATE)},
  pages={1047--1052},
  year={2022},
  organization={IEEE}
}

@article{ma2022survey,
  title={Survey on chiplets: interface, interconnect and integration methodology},
  author={Ma, Xiaohan and Wang, Ying and Wang, Yujie and Cai, Xuyi and Han, Yinhe},
  journal={CCF Transactions on High Performance Computing},
  volume={4},
  number={1},
  pages={43--52},
  year={2022},
  publisher={Springer}
}

@inproceedings{sahoo2024trade,
  title={Trade-off Between Chiplet Dimensions and Packaging Parameters for Optimal Performance to Cost for Chiplet Based Heterogeneous Integration},
  author={Sahoo, Krutikesh and Zhai, Max and Iyer, Subramanian},
  booktitle={2024 IEEE 26th Electronics Packaging Technology Conference (EPTC)},
  pages={232--237},
  year={2024},
  organization={IEEE}
}

@article{ma2023review,
  title={Review of wafer surface defect detection methods},
  author={Ma, Jianhong and Zhang, Tao and Yang, Cong and Cao, Yangjie and Xie, Lipeng and Tian, Hui and Li, Xuexiang},
  journal={Electronics},
  volume={12},
  number={8},
  pages={1787},
  year={2023},
  publisher={MDPI}
}

@inproceedings{shamshiri2009yield,
  title={Yield and cost analysis of a reliable NoC},
  author={Shamshiri, Saeed and Cheng, Kwang-Ting},
  booktitle={2009 27th IEEE VLSI Test Symposium},
  pages={173--178},
  year={2009},
  organization={IEEE}
}

@inproceedings{shamshiri2008cost,
  title={A cost analysis framework for multi-core systems with spares},
  author={Shamshiri, Saeed and Lisherness, Peter and Pan, Sung-Jui and Cheng, Kwang-Ting},
  booktitle={2008 IEEE International Test Conference},
  pages={1--8},
  year={2008},
  organization={IEEE}
}

@inproceedings{huang2010characterizing,
  title={Characterizing the lifetime reliability of manycore processors with core-level redundancy},
  author={Huang, Lin and Xu, Qiang},
  booktitle={2010 IEEE/ACM International Conference on Computer-Aided Design (ICCAD)},
  pages={680--685},
  year={2010},
  organization={IEEE}
}

@inproceedings{agrawal2023level,
  title={Level 4 Autonomous Driving SoC, leveraging chiplet, advanced package and UCIe},
  author={Agrawal, Vinayak and Piednoel, Francois and Elkanovich, Igor and Sil, Dwaipayan and Jahan, Mirza},
  booktitle={2023 IEEE Symposium on High-Performance Interconnects (HOTI)},
  pages={9--14},
  year={2023},
  organization={IEEE}
}

@article{cunningham1990use,
  title={The use and evaluation of yield models in integrated circuit manufacturing},
  author={Cunningham, James A},
  journal={IEEE Transactions on Semiconductor Manufacturing},
  volume={3},
  number={2},
  pages={60--71},
  year={1990},
  publisher={IEEE}
}

@inproceedings{lynch1977reduction,
  title={The reduction of LSI chip costs by optimizing the alignment yields},
  author={Lynch, WT},
  booktitle={1977 International Electron Devices Meeting},
  pages={7--7},
  year={1977},
  organization={IEEE}
}

@article{gwennap1993estimating,
  title={Estimating IC manufacturing costs: die size, process type are key factors in microprocessor cost},
  author={Gwennap, Linley},
  journal={Microprocessor Report},
  year={1993}
}

@inproceedings{mercier2006yield,
  title={Yield and cost modeling for 3D chip stack technologies},
  author={Mercier, P and Singh, SR and Iniewski, K and Moore, B and O'Shea, P},
  booktitle={IEEE Custom Integrated Circuits Conference 2006},
  pages={357--360},
  year={2006},
  organization={IEEE}
}

@article{tang2022costaware,
  title={Cost-Aware Exploration for Chiplet-Based Architecture with Advanced Packaging Technologies},
  author={Tang, Tianqi and Xie, Yuan},
  journal={arXiv preprint arXiv:2206.07308},
  year={2022}
}

@article{gopalakrishnan2011process,
  title={Process costing of the microchip},
  author={Gopalakrishnan, Bhaskaran and Gajera, Dipesh and Gupta, Deepak P and Athinarayanan, Ragu and Chaudhari, Subodh A},
  journal={International Journal of Industrial and Systems Engineering},
  volume={8},
  number={3},
  pages={326--345},
  year={2011},
  publisher={Inderscience Publishers}
}

@article{lauterbach2021path,
  title={The path to successful wafer-scale integration: The Cerebras story},
  author={Lauterbach, Gary},
  journal={IEEE Micro},
  volume={41},
  number={6},
  pages={52--57},
  year={2021},
  publisher={IEEE}
}

@inproceedings{xu2010tsv,
  title={A study of Through Silicon Via impact to 3D Network-on-Chip design},
  author={Xu, Canhao and Liljeberg, Pasi and Tenhunen, Hannu},
  booktitle={2010 International Conference on Electronics and Information Engineering},
  volume={1},
  pages={V1-370--V1-374},
  year={2010},
  organization={IEEE}
}

@article{chen2023floorplet,
  title={Floorplet: Performance-aware floorplan framework for chiplet integration},
  author={Chen, Shixin and Li, Shanyi and Zhuang, Zhen and Zheng, Su and Liang, Zheng and Ho, Tsung-Yi and Yu, Bei and Sangiovanni-Vincentelli, Alberto L},
  journal={IEEE Transactions on Computer-Aided Design of Integrated Circuits and Systems},
  volume={43},
  number={6},
  pages={1638--1649},
  year={2023},
  publisher={IEEE}
}

@inproceedings{ning2023supply,
  title={Supply chain aware computer architecture},
  author={Ning, August and Tziantzioulis, Georgios and Wentzlaff, David},
  booktitle={Proceedings of the 50th Annual International Symposium on Computer Architecture},
  pages={1--15},
  year={2023}
}

@article{roman2025ppac,
  title={PPAC Driven Multi-die and Multi-technology Floorplanning},
  author={Roman-Vicharra, Cristhian and Chen, Yiran and Hu, Jiang},
  journal={arXiv preprint arXiv:2502.10932},
  year={2025}
}

@inproceedings{hao2023monad,
  title={Monad: Towards cost-effective specialization for chiplet-based spatial accelerators},
  author={Hao, Xiaochen and Ding, Zijian and Yin, Jieming and Wang, Yuan and Liang, Yun},
  booktitle={2023 IEEE/ACM International Conference on Computer Aided Design (ICCAD)},
  pages={1--9},
  year={2023},
  organization={IEEE}
}

@inproceedings{yang2024challenges,
  title={Challenges and opportunities to enable large-scale computing via heterogeneous chiplets},
  author={Yang, Zhuoping and Ji, Shixin and Chen, Xingzhen and Zhuang, Jinming and Zhang, Weifeng and Jani, Dharmesh and Zhou, Peipei},
  booktitle={2024 29th Asia and South Pacific Design Automation Conference (ASP-DAC)},
  pages={765--770},
  year={2024},
  organization={IEEE}
}

@inproceedings{mallya2025performance,
  title={A Performance Analysis of Chiplet-Based Systems},
  author={Mallya, Neethu Bal and Strikos, Panagiotis and Goel, Bhavishya and Ejaz, Ahsen and Sourdis, Ioannis},
  booktitle={2025 Design, Automation \& Test in Europe Conference (DATE)},
  pages={1--7},
  year={2025},
  organization={IEEE}
}

@article{graening2025catch,
  title={CATCH: a Cost Analysis Tool for Co-optimization of chiplet-based Heterogeneous systems},
  author={Graening, Alexander and Talukdar, Jonti and Pal, Saptadeep and Chakrabarty, Krishnendu and Gupta, Puneet},
  journal={arXiv preprint arXiv:2503.15753},
  year={2025}
}

@article{graening2025chipletpart,
  title={ChipletPart: Scalable Cost-Aware Partitioning for 2.5 D Systems},
  author={Graening, Alexander and Gupta, Puneet and Kahng, Andrew B and Pramanik, Bodhisatta and Wang, Zhiang},
  journal={arXiv preprint arXiv:2507.19819},
  year={2025}
}

@inproceedings{ayes2025network,
  title={Network-on-Interposer Co-Design for Heterogeneous Chiplet-Based Integrated Systems},
  author={Ayes, Andres and Friedman, Eby G and Wolf, Marilyn},
  booktitle={2025 IEEE International Symposium on Circuits and Systems (ISCAS)},
  pages={1--5},
  year={2025},
  organization={IEEE}
}

@article{mishty2024ai,
  title={AI-aided System and Design Technology Co-optimization Methodology Towards Designing Energy-efficient and High-performance AI Accelerators},
  author={Mishty, Kaniz Fatema and others},
  year={2024}
}

@misc{UCIe22,
  title        = {{Universal Chiplet Interconnect Express (UCIe) Specification}},
  author       = {{UCIe Consortium}},
  year         = {2022},
  note         = {Version 1.0},
  howpublished = {\url{https://www.uciexpress.org}}
}

@misc{AIB18,
  title        = {{Advanced Interface Bus (AIB) Specification}},
  author       = {{Intel Corporation}},
  year         = {2018},
  howpublished = {\url{https://www.intel.com}}
}

\vspace{-10 mm} 
\begin{IEEEbiography}
[{\includegraphics[width=1in,clip,keepaspectratio]{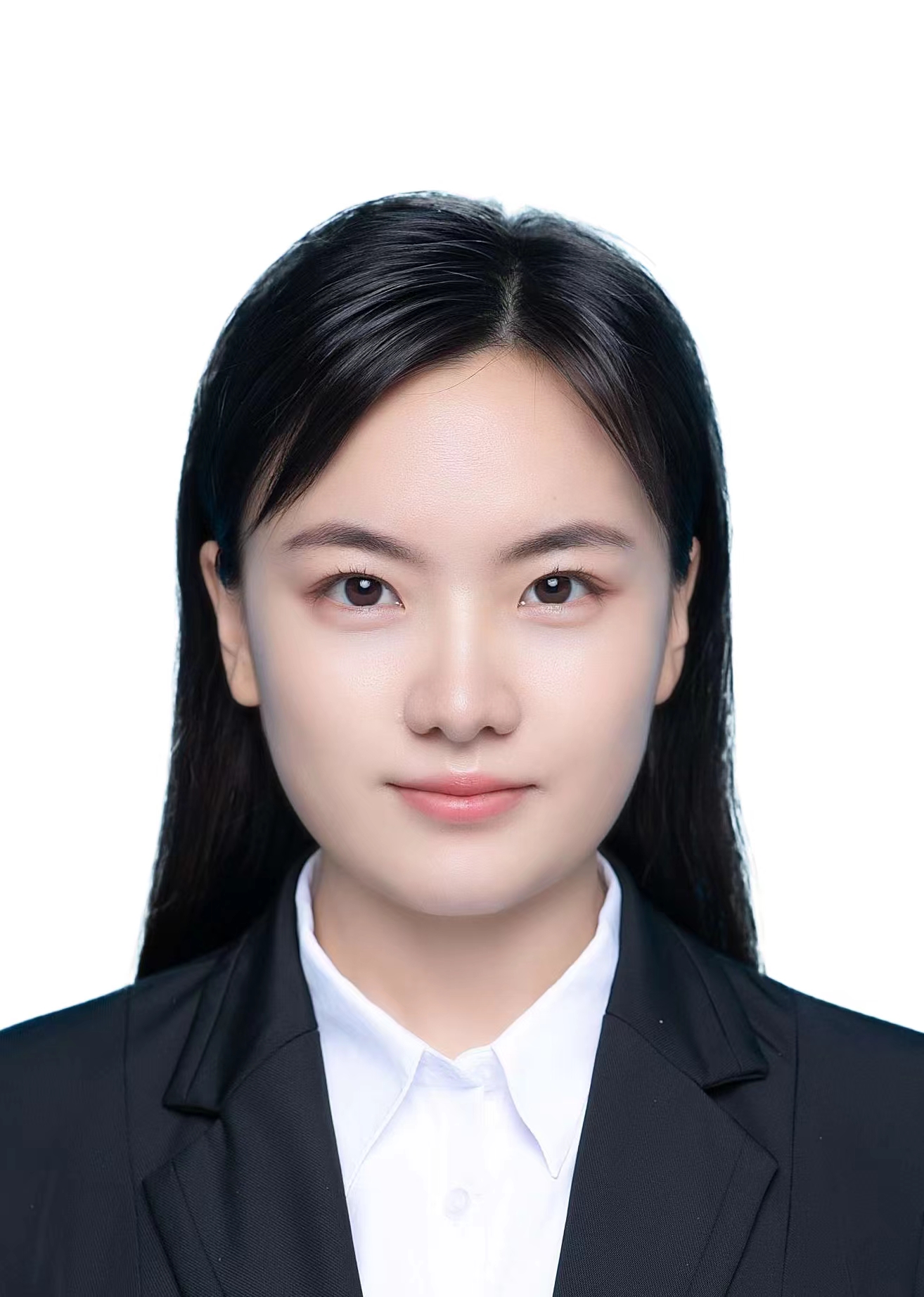}}]{Zizhen Liu} (Member, IEEE) 
received the B.S. degree in Hefei University of Technology,
Hefei, China, in 2018, and the
the Ph.D. degree from the Institute of
Computing Technology (ICT), Chinese Academy of
Sciences (CAS), Beijing, China, in 2023. She is currently an Assistant Professor with the
State Key Laboratory of Processors, ICT, CAS.
Her research interests include fault-tolerant computing, design automation, and AI security.
\vspace{-10 mm} 
\end{IEEEbiography}

\begin{IEEEbiography}
[{\includegraphics[width=1in,clip,keepaspectratio]{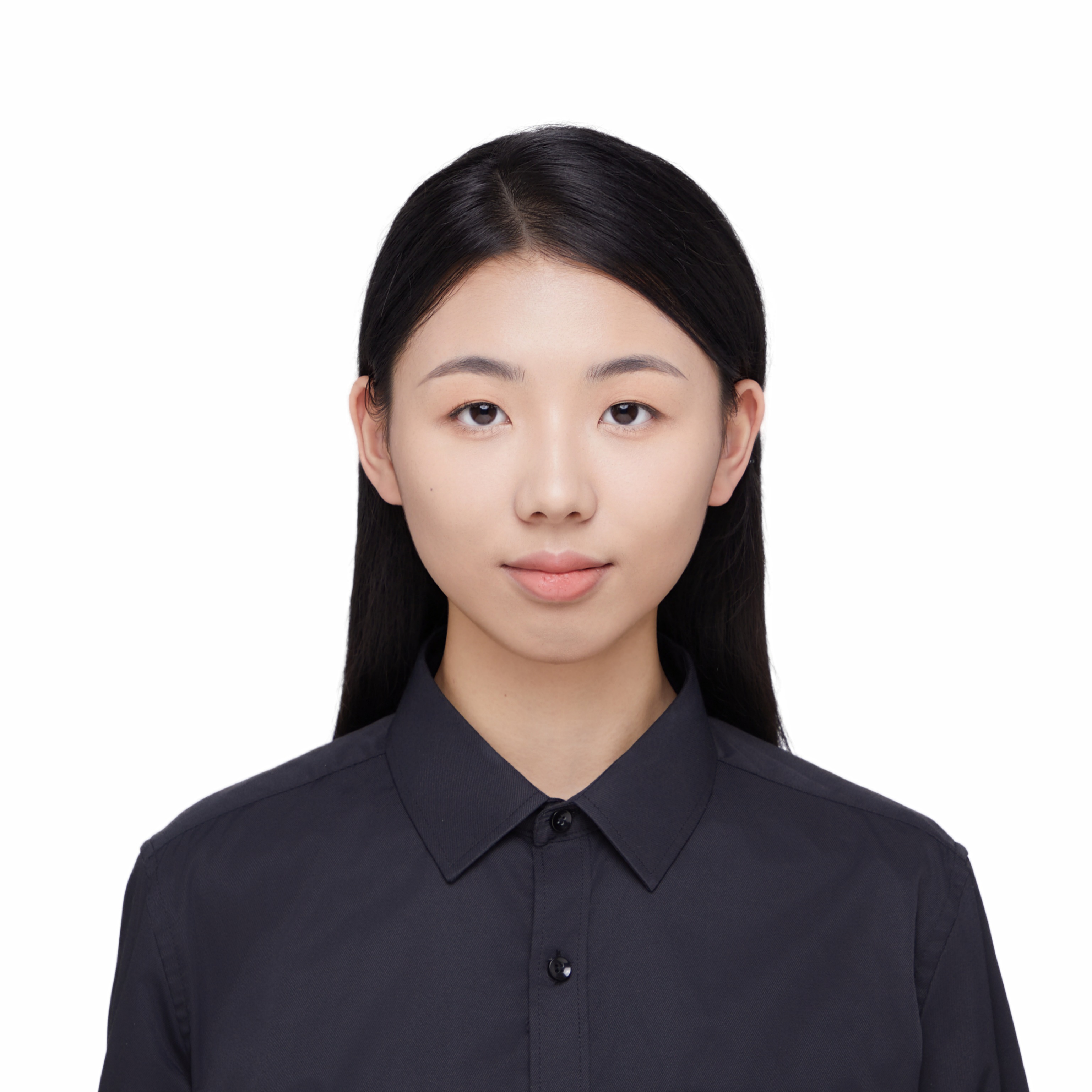}}]{Fangzhiyi Wang} received the B.S. degree from Guangxi University in 2025. She is currently pursuing the Ph.D. degree at the School of Advanced Interdisciplinary Sciences, University of Chinese Academy of Science University. Her research interests include fault-tolerant design and computer architecture.
\vspace{-13 mm} 
\end{IEEEbiography}

\begin{IEEEbiography}
[{\includegraphics[width=1in,clip, keepaspectratio]{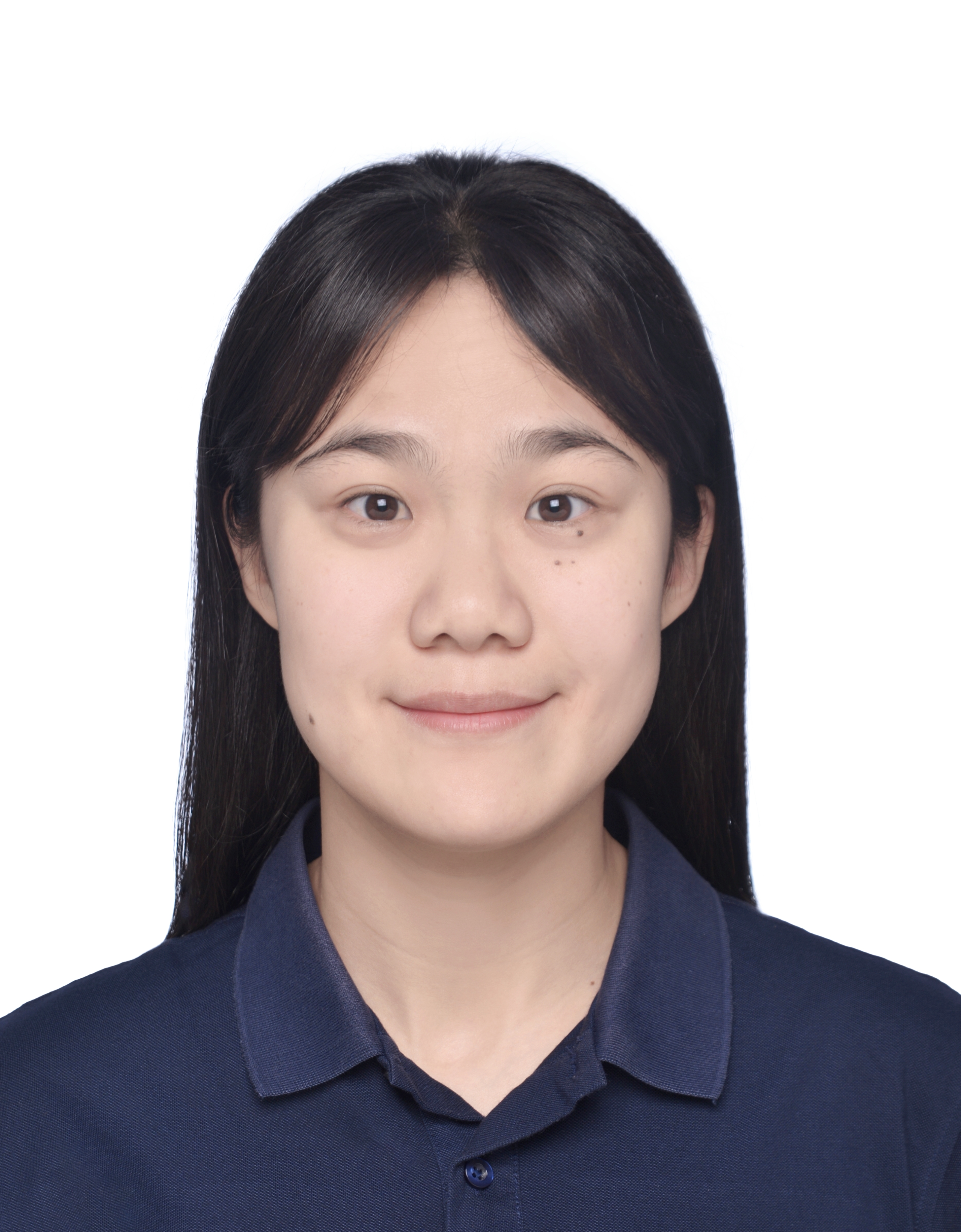}}]{Mengdi Wang}
(Member, IEEE) received the BE degree in computer science from Huazhong University, Wuhan, China, in 2017, and the PhD degree in computer science from the Institute of Computing Technolog (ICT), Chinese Academy of Sciences (CAS), Beijing, in 2023. She is currently an assistant researcher with ICT, CAS. Her research interests include computer architecture and VLSI.
\vspace{-13 mm} 
\end{IEEEbiography}

\begin{IEEEbiography}
[{\includegraphics[width=1in,clip,keepaspectratio]{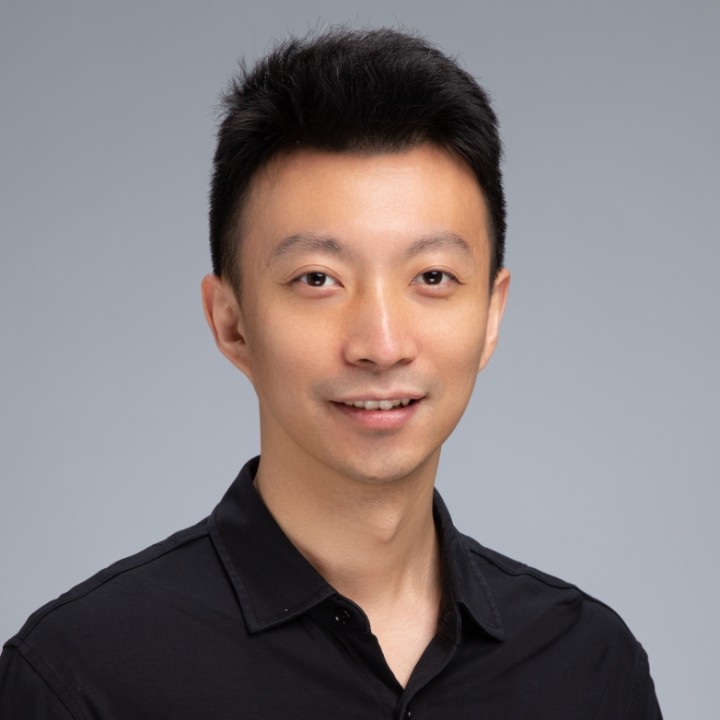}}]{Jing Ye} is currently an associate professor of State
Key Laboratory of Processors (SKLP), Institute of
Computing Technology (ICT), Chinese Academy of
Sciences (CAS), and the CEO of CASTEST corporation. He received the Ph.D. degree in ICT, CAS in
2014, and the B.S. degree in Electronics Engineering
and Computer Science, Peking University in 2008.
His current research interests include VLSI test,
hardware security, cryptographic, PUF, PQC, and AI
software/hardware security.
\vspace{-12 mm} 

\end{IEEEbiography}

\begin{IEEEbiography}
[{\includegraphics[width=1in,clip,keepaspectratio]{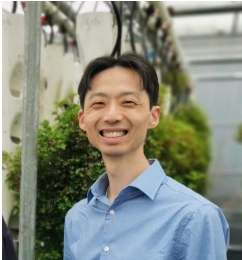}}]{Hayden Kwok-Hay So} (Senior Member, IEEE) received the B.S., M.S., and Ph.D. degrees in electrical engineering and computer sciencesm from the University of California, Berkeley, CA, USA, in 1998, 2000, and 2007, respectively. He is the Director of the School of Innovation and the co-director of the Joint Lab on Future Cities, The University of Hong Kong (HKU).  He is a Croucher Innovation Awardee (2013) and a recipient of the CODES+ISSS Test-of-Time Award for foundational work in hardware–software co-design.  He leads the Computer Architecture \& System Research Lab, where his team develops reconfigurable computing platforms and hardware–software co-design methodologies spanning FPGA architectures, operating systems, and high-performance applications.
\vspace{-13 mm} 
\end{IEEEbiography}

\begin{IEEEbiography}
[{\includegraphics[width=1in,clip,keepaspectratio]{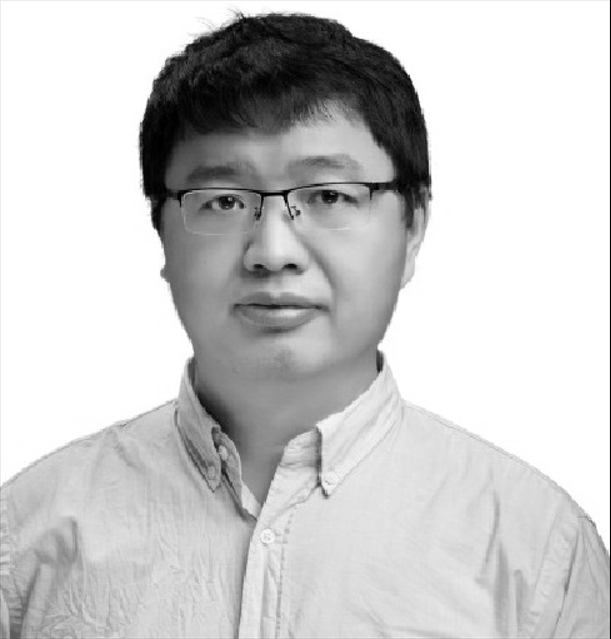}}]{Cheng Liu} (Senior Member, IEEE)  received the B.Eng. and M.Eng. degrees from Harbin Institute of Technology, Harbin, China, in 2007 and 2009, respectively, and the Ph.D. degree from The University of Hong Kong, Hong Kong, in 2016. He is currently an Associate Professor with Institute of Computing Technology, Chinese Academy of Sciences, Beijing, China. His research interests include domain specific architectures (DSA), DSA design automation, and fault-tolerant computing.
\vspace{-13 mm} 
\end{IEEEbiography}

\begin{IEEEbiography}
[{\includegraphics[width=1in,clip,keepaspectratio]{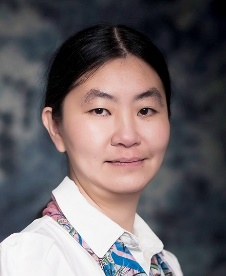}}]{Huawei Li} (M’00–SM’09) received the B.S. degree in computer science from Xiangtan University,
Xiangtan, China, in 1996, and the M.S. and Ph.D.
degrees from the Institute of Computing Technology
(ICT), Chinese Academy of Sciences (CAS), Beijing, China, in 1999 and 2001, respectively. She has
been a Professor with ICT and CAS since 2008. She
was a visiting Professor at the University of California at Santa Barbara, Santa Barbara, CA, USA, from
2009 to 2010. 

She currently serves as the Secretary
General of the China Computer Federation (CCF)
Technical Committee on Integrated Circuit Design, the Steering Committee
Chair of IEEE Asian Test Symposium (ATS), and the Associate Editor of
IEEE Transactions on Very Large Scale Integration (VLSI) Systems and
IEEE Design Test. Her current research interests include testing of very
large-scale integration/SoC circuits, approximate computing architecture and
machine learning accelerators. She has published more than 200 technical
papers and holds 34 Chinese patents in the above areas.

\end{IEEEbiography}
\end{document}